\documentclass[%
  twocolumn,%
  unsortedaddress,%
  showpacs,%
  letterpaper,%
  amsfonts,%
  prb,%
  bibnotes,%
  floatfix%
]{revtex4}

\usepackage[nolist,nohyperlinks]{acronym}
\usepackage{subfigure,graphicx}
\usepackage{amsmath,amssymb,amsxtra,bm,bbm}
\usepackage{hyperref}
  \hypersetup{%
    breaklinks = {true},
    citecolor = {blue},
    colorlinks = {true},
    linkcolor = {red},
    pdfauthor = {\textcopyright\ Vitor M. Pereira},
    pdfcreator = {\LaTeX\ and \flqq hyperref\frqq},
    pdffitwindow = {true},
    pdfmenubar = {true},
    pdfpagelayout = {SinglePage},
    pdfstartview = {Fit},
    pdftoolbar = {true},
    plainpages = {false},
  }

\newcommand{\abs}[1]{\left| #1 \right|}

\newcommand{\average}[1]{\left\langle #1 \right\rangle}

\newcommand{\Fcal}{\mathcal{F}}
\newcommand{\Tc}{\ensuremath{T_\text{C}}}
\newcommand{\Wp}{\ensuremath{\omega_\text{p}}}
\newcommand{\Si}{\ensuremath{\vec{S}_i\,}}
\newcommand{\Ef}{\ensuremath{E_\text{F}\,}}
\newcommand{\fref}[1]{Fig.~\ref{#1}}
\newcommand{\eref}[1]{Eq.~\eqref{#1}}
\newcommand{\sref}[1]{Sec.~\ref{#1}}
\newcommand{\Ref}{Ref.}

\DeclareMathOperator{\tr}{Tr}

\begin{document}


\title{
Double Exchange Model at Low Densities:
Magnetic Polarons and Coulomb Suppressed Phase Separation
}



\author{Vitor M. Pereira}
\affiliation{Department of Physics, Boston University, 590 
Commonwealth Avenue, Boston, MA 02215, USA}

\author{J. M. B. Lopes~dos~Santos}
\affiliation{CFP and Departamento de F{\'\i}sica, Faculdade de Ci\^encias
Universidade de Porto, 4169-007 Porto, Portugal}

\author{A.~H. Castro Neto}
\affiliation{Department of Physics, Boston University, 590 
Commonwealth Avenue, Boston, MA 02215, USA}

\date{\today}


\begin{abstract}
We consider the double exchange model at very low densities. The conditions for
the formation of self-trapped magnetic polarons are analyzed using an
independent polaron model. The issue of phase separation in the low density
region of the temperature-density phase diagram is discussed. 
We show how electrostatic and localization
effects can lead to the substantial suppression of the phase separated regime.
By examining connections between the resulting phase and the polaronic phase,
we conclude that they reflect essentially the same physical situation of a
ferromagnetic droplet containing one single electron. In the
ultra diluted regime, we explore the possible stabilization of a Wigner
crystal of magnetic polarons.
Our results are compared with the experimental evidence for a polaronic phase
in \acf{EuB6}, and we are able to reproduce the experimental region of stability
of the polaronic phase. We further demonstrate that phase-separation is a
general feature expected in metallic ferromagnets whose bandwidth depends on the
magnetization.
\end{abstract}

\maketitle


%
\section{Introduction}%
\label{sec:Intro}

The earliest descriptions of the concept of magnetic polaron appear with the
considerations by de~Gennes on the relevance of the \acf{DE} mechanism to the
mixed valent manganites \cite{Gennes:1960}, and then with Nagaev's studies of
antiferromagnetic semiconductors, who coined the term \emph{ferron} also often
used in the context \cite{Nagaev:2001,Nagaev:2002}. 
Some of the first theories based upon the presence of magnetic
polarons were developed later in the context of the ferromagnetic semiconductor
EuO, to explain the spectacular metal-insulator transition found at the onset of
ferromagnetism for the Eu-rich samples of this compound
\cite{Oliver:1972,Torrance:1972}. Further developments on this concept
permitted the explanation of the spin-flip Raman scattering characteristic of
certain \ac{DMS} \cite{Dietl:1982,Dietl:1983,Heiman:1983,Torrance:1972} and,
more recently, their presence was also claimed to be present in the \ac{EuB6}
hexaborides \cite{Snow:2001,Nyhus:1997}. 

Just as in the analogous case of the electrostatic polaron, one can devise a
pictorial description of a magnetic polaron in real space, consisting of a
charge carrier surrounded by a cloud of polarized local spins in an unpolarized
magnetic background --- a state that arises from the exchange interaction
between the carrier spin and the lattice spins. A distinction is usually made
between the so-called \ac{BMP} and the \ac{FMP}. The \ac{BMP} is invoked when
the charge carrier is bound via Coulomb interaction to an impurity center, and
is typical in the magnetic semiconductors. In this case the trapped carrier
polarizes the lattice spins within its effective Bohr radius as a consequence of
the \emph{s--d}-like interaction between electron and lattice spins. The
\ac{FMP}, by contrast, results from the fact that a free carrier interacting
with lattice spins via an \emph{s--d} coupling, can minimize its kinetic energy
by polarizing its vicinity. Under certain conditions this carrier can then
become self-trapped in the resulting potential well created by the effect of the
local ferromagnetism. 

The presence of such type of entities is believed to play an important role in
the emergence of many interesting properties of several important magnetic
materials: many peculiarities of the manganites, such as the \ac{CMR} effect and
other anomalies in its transport and magnetism, have been attributed in part to
the development of magnetic polarons near the ferromagnetic
transition (although in this case the influence of orbital and lattice
degrees of freedom is equally important)
\cite{Teresa:1997a,Amaral:1998,Teresa:2000}; an interpretation for the \ac{CMR}
in the Mn pyrochlores was also proposed on the basis of magnetic polarons
\cite{Majumdar:1998}; besides their relevance for the already mentioned
Eu-chalcogenides and II-VI semiconductors, in the III-V \ac{DMS} like
Ga$_{1-x}$Mn$_{x}$As, the ferromagnetic transition can be interpreted within a
\ac{BMP} percolating scenario \cite{Kaminski:2002}; in \ac{EuB6}, our target
magnetic metal exhibiting \ac{CMR}, their presence in signaled in the optical
response \cite{Nyhus:1997,Snow:2001}, although the theoretical interpretation
of these results has been subject to questioning \cite{Calderon:2004}. 

Given that these different classes of magnetic materials have attracted
considerable attention in recent years because of their potential for the
development of new magnetoelectronic devices, and since magnetic polarons have
an apparently ubiquitous presence among them, a number of theoretical approaches
to the problem have been developed through the years. In particular, extensive
work has been done with emphasis in the physics of magnetic semiconductors
\cite{Kasuya:1970,Dietl:1982,Heiman:1983,Mauger:1983,Nagaev:2002} and \ac{CMR}
manganites \cite{Varma:1996,Batista:2000,Garcia:2002,Meskine:2004}. 

In the ensuing sections we will focus our attention on the stability
conditions for the free magnetic polaron in the \ac{DEM}, with an eye on the
experimental evidence for magnetic polarons in \ac{EuB6}
\cite{Nyhus:1997,Snow:2001}, and having in mind the description of the
magneto-electronic properties of this material in terms of \ac{DE}
\cite{Pereira:2004}.
Studies devoted to the polaronic stability in this particular model and its
variations have been performed by several authors both analytically and
numerically, although under different assumptions
\cite{Daghofer:2004a,Daghofer:2004b,Garcia:2002,Koller:2003,Neuber:2005,
Pathak:2001,Wang:1997,Yi:2000}. We show that the \ac{DE}-based
interpretation of magnetotransport in \ac{EuB6} is consistent with the
experimental evidence for a polaronic phase mediating the \acs{PM}-\acs{FM}
transition. In particular, within an \ac{IPM} we reproduce the experimental
temperature and density range of the polaronic phases, without adjusting
parameters.

Given that we are focusing on the low density regime of the \ac{DEM}, another
issue becomes pressing: the known tendency for this model to exhibit a phase
separation instability at reduced densities
\cite{Nagaev:1998,Yunoki:1998,Arovas:1999,Kagan:1999,Alonso:2001a,Alonso:2001b}.
We characterize the phase-separated regime and study how the introduction of
electrostatic corrections leads to considerable shrinking of the
phase separation region in the phase diagram. We discuss how the
phase-separated regime is connected and compatible with the polaronic phase.

This paper is organized in the following manner. In \sref{sec:Overview-EuB6} we
briefly introduce some experimental details regarding electronic transport and
magnetism in \ac{EuB6}, only to the extent of motivating our studies. In
\sref{sec:DEM-MagPol-EuB6} we introduce and discuss an \ac{IPM} for the
\ac{DEM} at very low densities, whose results are then confronted in
\sref{sec:DEM-MagPol-EuB6} with the experimental evidence for magnetic polarons
in \ac{EuB6}. We then address the problem of phase separation in the
low-density \ac{DEM} in \sref{sec:DEM-PhSeparation}: we discuss the emergence
of phase separation at low densities, study its suppression when
electrostatic corrections are taken into consideration and explore the
connections of the resulting phase with the polaronic phase. We consider the
ultra diluted regime in \sref{sec:WignerCrystal}, and provide estimates for the
stability of a Wigner crystal of magnetic polarons. In \sref{sec:Conclusions} we
close this paper with our conclusions.

%
\begin{figure}
  \centering
  \subfigure[][]{
    \includegraphics*[width=0.26\textwidth]{%
      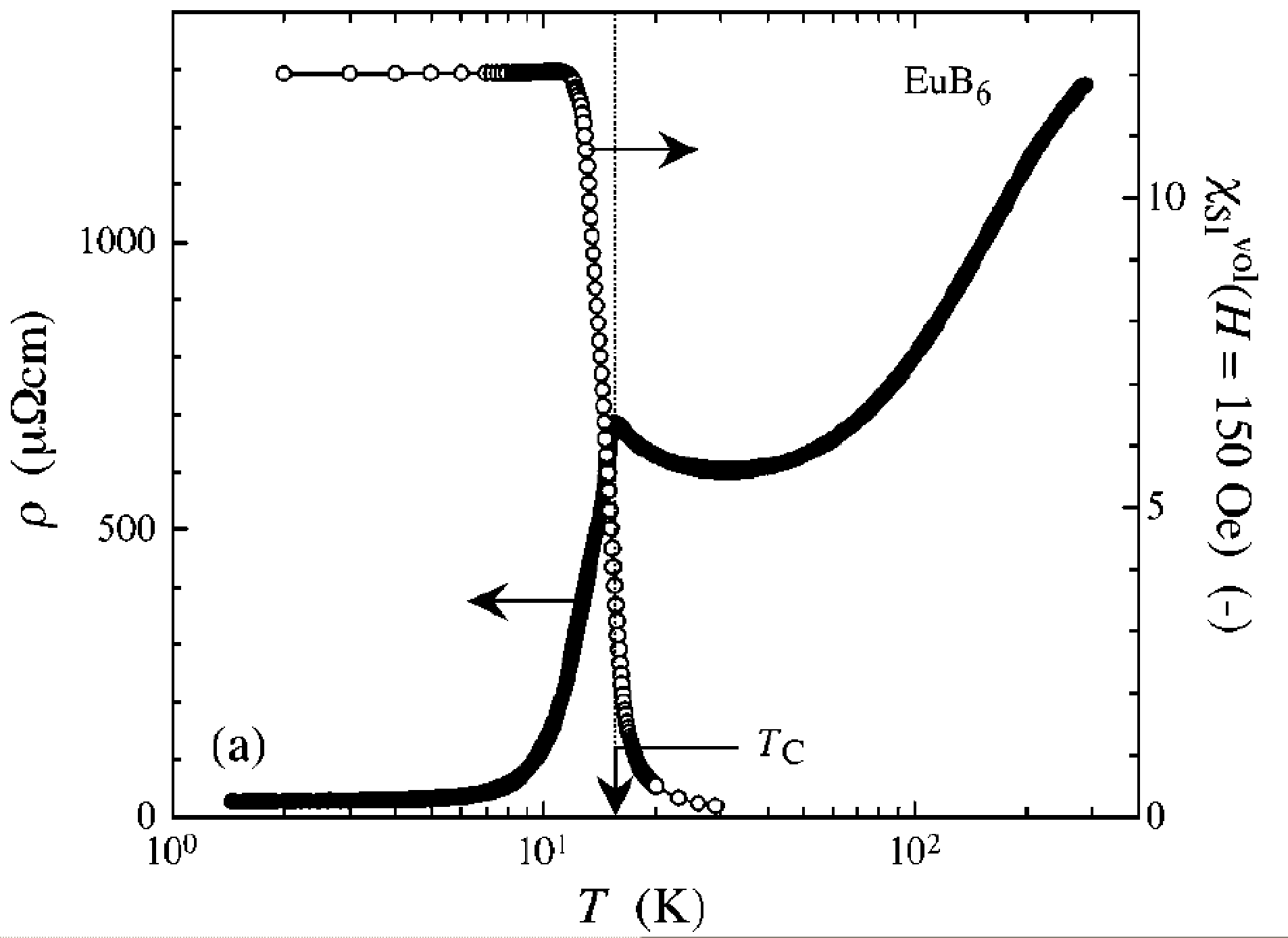}%
    \label{fig:EuB6-Exp-Transport}%
  }%
  \subfigure[][]{
    \includegraphics*[width=0.20\textwidth]{%
      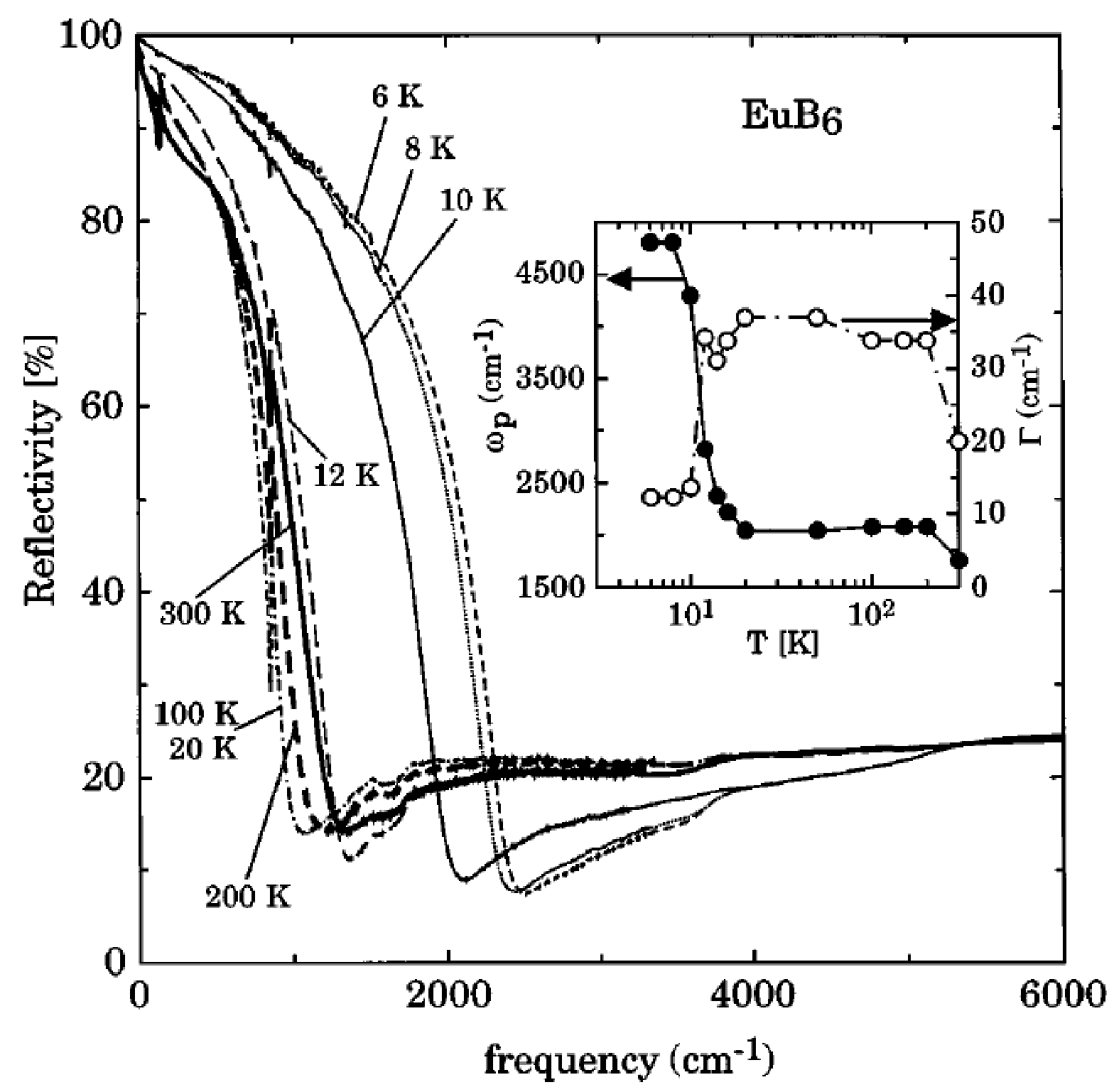}%
    \label{fig:EuB6-Exp-Plasma}%
  }
  \caption{
    Some experimental signatures of \ac{EuB6}. 
    Panel \subref{fig:EuB6-Exp-Transport} shows the behavior of the DC
    resistivity and magnetic susceptibility as a function of temperature, as
    reproduced in \Ref~\onlinecite{Paschen:2000}.
    Notice the upturn followed by a massive drop in $\rho(T)$ at \Tc, which 
    correlates with the onset of magnetic order.
    In \subref{fig:EuB6-Exp-Plasma} we reproduce the results from
    \Ref~\onlinecite{Degiorgi:1997} for the optical reflectivity at different
    temperatures. The enhancement of \Wp\ as the system becomes magnetic 
    is clear.
  }
  \label{fig:EuB6-Experiments}
\end{figure}
%

%
\section{Overview of EuB$_\text{6}$ Properties}%
\label{sec:Overview-EuB6}

We briefly review the physics of \ac{EuB6} that motivates our choice of
microscopic model. An extensive review of the phenomenology of this material
was given in \Ref~\onlinecite{Pereira:2006PhD}, and here we limit ourselves to
the essential aspects relevant in the context of this paper.
\ac{EuB6} is a magnetic metal\cite{Fisk:1979,Guy:1980}, with residual
resistivities at $T\to 0$ of the order of $\sim10\mu\Omega.\text{cm}$, or less,
and exibiths a clear \ac{CMR} signal close to \Tc
\cite{Sullow:1998,Aronson:1999} [\fref{fig:EuB6-Exp-Transport}].
Notwithstanding, \ac{EuB6} has a very small carrier density. More precisely,
carrier densities estimated from Hall effect measurements amount to as little as
$\sim 10^{19}\text{cm}^{-3}$, or $\sim 0.001$ carriers per unit cell
\cite{Fisk:1979,Paschen:2000,Rhyee:2003c}. 
These conduction electrons are believed to arise from the presence of defects in
the structural arrangement of the boron framework
\cite{Blomberg:1995,Monnier:2001}. Such defects generate a surplus of electrons
that occupy states in the conduction band.

One of the most intriguing and peculiar features of \ac{EuB6} is arguably the
giant and rather unusual blue-shift of the unscreened plasma edge, \Wp, induced
simply by a temperature variation \citep{Broderick:2003,Degiorgi:1997}
[\fref{fig:EuB6-Exp-Plasma}]. 
At $T>\Tc$, the reflectivity spectrum displays a typical metallic behavior, with
a very well defined plasma threshold. With the establishment of the long-range
magnetic order, the plasma edge increases markedly in such a way that \Wp varies
by a factor of almost 3 between \Tc\ and $T\ll\Tc$
\citep{Broderick:2002,Degiorgi:1997}. This variation of \Wp\ is consistent with
the remarkable enhancement in the carrier densities as the
temperature is lowered past \Tc \cite{Paschen:2000}.

The conduction and valence bands of \ac{EuB6} are separated by a gap of the
order of 1~eV \cite{Denlinger:2002,Gianno:2002,Souma:2003,Rhyee:2003a}. This
fundamental gap lies at the $X$ point in the cubic \ac{BZ}, and the close
proximity of \Ef to the bottom of the conduction band dictates a pocket-like
structure for the Fermi surface. Given that the interest will be almost
completely in a temperature range $T\lesssim\Tc$, the electronic
states in the valence band are disregarded in the following.

The magnetism of \ac{EuB6} arises entirely from the half-filled $4f$
shell of Eu$^{2+}$ in the state $^{8}S_{7/2}$. This implies localized magnetism
stemming from magnetic moments of magnitude $S=7/2$. Within this formulation
these electrons do not itinerate at all. We designate the resulting magnetic
moment by \emph{local spin}, and use the term \emph{magnetization of the system}
when alluding to long range ordered phases of these spins. In addition, for our
purposes the conduction band electrons interact with the local spins only
through the Hund's coupling between the electron's spin and the local
moments'.

%
\section{The Independent Polaron Model}%
\label{sec:DEM-MagPol-IPM}

The Hamiltonian describing conduction electrons hopping in a tridimensional
cubic lattice, and coupled to local spins at each lattice site is
\begin{align}
  \mathcal{H}_{\,_{KLM}} &= \sum_{\left<i,j\right>, \sigma} 
        t_{i,j} \; c^\dag_{i,\sigma} c_{j,\sigma} + \text{h.c.} 
  \nonumber\\
    & + J_H \sum_{i,\alpha,\beta} 
        \Si \cdot \vec{\tau}_{\alpha,\beta} \; c^\dag_{i,\alpha} c_{i,\beta}
  \label{eq:EuB6-Hamiltonian-KLM}
  \,,
\end{align}
usually known as $s-f$ or \ac{KLM} Hamiltonian \citep{Doniach:1977}, and is one
of the canonical models in correlated electronic systems. In this expression,
$t_{i,j}\equiv t$ is the hopping integral between neighboring lattice sites,
$c^\dag_{i,\sigma} (c_{i,\sigma})$ are the second-quantized fermionic creation
(annihilation) operators at latice site $i$, \Si represents the local magnetic
moment of magnitude $S=7/2$, $J_H$ the exchange coupling of the latter to the
itinerating electrons, and $\vec{\tau}=(\tau_1,\tau_2,\tau_3)$ is the vector of
Pauli matrices. The sum in the first term is over all pairs of nearest
neighboring sites $\left<i,j\right>$.

Since $S=7/2$ is a rather high spin, the local spin operator is replaced by the
classical vector \Si, parametrized with spherical angles as $\Si = S \bigl(
\sin(\theta_i)\cos(\varphi_i), \sin(\theta_i)\sin(\varphi_i), \cos(\theta_i)
\bigr)$. This transforms the second term in \eqref{eq:EuB6-Hamiltonian-KLM} into
a generalized potential term for the electrons, which will be a disordered
potential above \Tc, where all \Si are uncorrelated. Furthermore, the magnetic
and electronic time scales are well apart in such a way that the magnetic
background provided by the \Si is essentially quenched. This means that the
typical time between spin fluctuations is much longer than the time for the
electronic subsystem to reach its ground state%

%
\begin{figure}
  \centering
  \includegraphics*[width=0.8\columnwidth]{%
    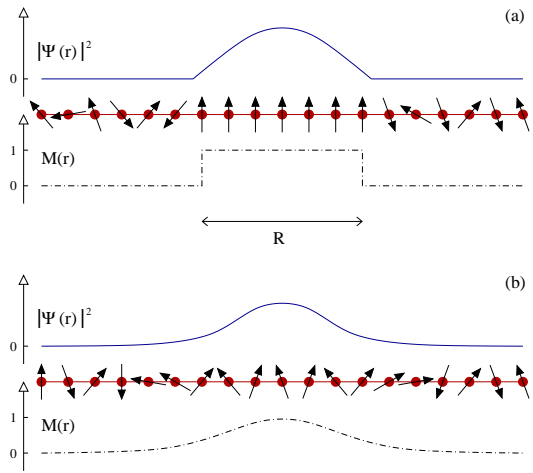}
  \caption{
    Contrasting the variational approach considered in
    eq.~\eqref{eq:MagPol-DF-WithConfEntropy} (a) to the more realistic,
    self-consistent, situation (b). Represented are the wave function of the
    self-localized electron, a possible spin configuration and the magnetization
    profile.
  }
  \label{fig:MagPol-Polaron-Picture}
\end{figure}
%

Under these circumstances, we consider the effective \ac{DE} Hamiltonian
\cite{Gennes:1960,Hasegawa:1979} that obtains in the limit $J_H\to\infty$. This
is accomplished through a local rotation of the quantization axis
so that it coincides with the direction of \Si at each site, and projecting out
the anti-parallel electron states \cite{Dagotto_Book:2003}. Such anti-parallel
states lie higher in energy (by $\approx J_H$) and hence are suppressed in the
effective Hilbert space. The result is 
\begin{equation}
  \mathcal{H}_{DE} = t \sum_{\langle ij \rangle} a_{ij} d^\dagger_i d_j
    + \text{H. c.}
  \,,
  \label{eq:DEM-DE-Hamiltonian-2}
\end{equation}
where the new operators $d_i$ correspond to an effective spinless electron that
maintains its spin aligned with each local moment, and all information about
the magnetic background is condensed in the effective hopping amplitude
$a_{ij}$:
\begin{equation}
  a_{ij} = \cos\left(\frac{\theta_i}{2}\right)
            \cos\left(\frac{\theta_j}{2}\right)
         + \sin\left(\frac{\theta_i}{2}\right)
            \sin\left(\frac{\theta_j}{2}\right)
           e^{-i(\phi_{i}-\phi_{j})}
  \,.
\end{equation}
Within the simple \ac{DEM}, where no other polaron-favoring interactions are
included (e.g. not including \ac{AFM} exchange terms between the local
moments.), the only possibility for the stabilization of magnetic polarons is at
low electronic densities. This happens because the local spins of several
neighboring unit cells are expected to participate in the magnetization cloud of
each electron. Were it otherwise (i.e. at high electronic densities), the
electronic wave functions would overlap considerably destroying this polaron
picture. 

In order to investigate this problem we will first discuss the
thermodynamic stability of magnetic polarons within the \ac{DEM}. We can write a
free energy for the system including an electronic contribution consisting of
the ground state energy for the electrons in a given magnetic configuration, the
local spins contributing only with an entropic term. To tackle the first
part, we calculate numerically the exact electronic \ac{DOS} for a given
spin configuration \cite{Pereira:2006PhD}, which we designate by
$\rho\bigl(E,\{\Si\}\bigr)$, and extract the disorder-averaged \ac{DOS} at
constant magnetization: $\rho\bigl(E,M)$ ($M$ being the normalized
local magnetization) (details about this numerical calculation will be given
below in \sref{sec:DEM-Disorder-Thermodynamics}). 

With this averaging over disorder, we can write the electronic contribution
to the free energy as 
\begin{equation}
  E_{el}(M,n_{e}) = \int\Theta\Bigl(E_{F}(M,n_{e})-\epsilon\Bigr) E
                    \rho(E,M) dE
  \,,
  \label{eq:MagPol-Eel_Full}
\end{equation}
where the dependence of the Fermi energy on both magnetization and electron
density was made explicit. When $n_{e}\ll1$, and for the purposes of the current
calculation, \eref{eq:MagPol-Eel_Full} can be approximated simply by
\begin{equation}
  E_{el}(M,n_{e}) \approx E_{b}(M) n_{e}
  \label{eq:MagPol-Eel_Bottom}
  \,,
\end{equation}
with $E_{b}$ representing the bottom of the band -- it becomes just a single
electron problem. Indeed, given the nature of the calculation and approximations
involved here, the consideration of the finite band filling introduces only
minor corrections and thus we proceed with the above approximation to the
electronic energy (Appendix~\ref{Appendix:MagPol-FiniteBandFilling}). 
Now, for a disordered system, the concept of \emph{``bottom of the band''} has
to be taken carefully as the \ac{DOS} will always exhibit Lifshitz exponential
tails. In this case, from the calculations of the averaged \ac{DOS}, the
\emph{bottom} of the band is found to lie at $-4t$ for the \ac{PM} ($M=0$) case
\cite{Endnote-2} (cfr. \fref{fig:DEM-DOS-DE-Demo}). 
We intend to construct the polaronic phase having the paramagnetic, uniform,
phase as reference. 
Within a virtual-crystal approximation, the electron at the bottom of the band
has an energy of $-4t$, and will be extended throughout the system. 
If a region of ferromagnetism develops locally, its reference energy in this
region will be lowered to $-6t$, but there is an extra energy that has to be
paid if the electron is to become confined to this region. 
For simplicity let us assume that the polaron so formed consists of a region
inside a cube of side $R$ (in units of the lattice parameter, $a$), inside which
$M=1$.
Obviously, given that in the \ac{DEM} the magnetic interaction is mediated by
electron itinerancy, one expects that, once the electron localizes inside this
cube, there will be no magnetism outside, at any temperature. 
The variation of free energy per lattice site when going from the \ac{PM}
homogeneous phase to this polaronic one will be written as
\begin{widetext}
\begin{equation}
  \Delta \Fcal_{\textrm{Pol}}(R,T) = 
    4 t n_e - 6 t n_e \cos\left(\frac{\pi}{R+1}\right) +
    T n_e R^3 \log(2S+1) - 
    T\mathcal{S}_{\textrm{Cfg}}(n_e,R)
  \label{eq:MagPol-DF-WithConfEntropy}
  \,, 
\end{equation}
\end{widetext}
and reflects the two competing effects at play: the first is the electron's
preference for a ferromagnetic background accompanied by an energy cost for the
localization; the second is the reduction of entropy caused by the appearance of
the (fully polarized) magnetic polarons. The last contribution,
$\mathcal{S}_{\textrm{Cfg}}$, expresses a configurational entropy, related to
the spacial distribution of the polarons inside the system. It is a
combinatorial term that can be approximated by 
\begin{equation}
  \mathcal{S}_{\textrm{Cfg}} \simeq \frac{1}{R^3} \log\left(\frac{1}{1 - n_e
R^3}\right) - 
    n_e \log\left(\frac{n_e R^3}{1 - n_e R^3}\right)  
  \label{eq:MagPol-S_Config}
  \,,
\end{equation}
but, since we are working with $n_{e}\ll1$ and the polaronic system is below the
percolation threshold, it happens to be the smallest contribution to
$\Delta\mathcal{S}$, allowing us to neglect it without important quantitative
consequences. 

In writing eq.~\eqref{eq:MagPol-DF-WithConfEntropy} some important assumptions
were made regarding the wave function of the electron. Reasoning in terms of the
original band state of the electron, it is clear that when the magnetic
background polarizes, the electron energy is lowered by $2t$. Thus, the
potential well is, at most, $2t$ deep and any bound state will always exhibit
exponential leaking of the wave function to the outside, whereas the energy of
the bound state in \eqref{eq:MagPol-DF-WithConfEntropy} was chosen as the energy
of an electron inside an infinite potential well. On the other hand, since one
can think of an effective magnetic coupling as proportional to
$\left|\psi_{e}(\vec{r})\right|^{2}$, the magnetization profile of the polaron
should also display a smooth variation, whereas in
eq.~\eqref{eq:MagPol-DF-WithConfEntropy} $M=0$ outside and $M=1$ inside the
polaron (cfr. \fref{fig:MagPol-Polaron-Picture}). These statements amount to say
that the exact treatment of the problem
requires a self-consistent
calculation of the bound state starting from the \ac{DEM} or perhaps from the
full Hamiltonian \cite{Kasuya:1970,Pathak:2001}.
In this sense, eq.~\eqref{eq:MagPol-DF-WithConfEntropy} is to be understood in
the spirit of a variational approach, $R$ being the variational parameter.

%
\begin{figure}[t]
  \centering
  \subfigure[][]{
    \includegraphics*[width=0.45\columnwidth]{%
      Figs/DOS_Exact_vs_Lanc_2D_R-9_N-50}%
    \label{fig:MagPol-DOS-and-Spectrum-1}
  }\quad%
  \subfigure[][]{
    \includegraphics*[width=0.45\columnwidth]{%
      Figs/DOS_Exact_vs_Lanc_3D_R-5_N-50}%
    \label{fig:MagPol-DOS-and-Spectrum-2}
  }
  \subfigure[][]{
    \includegraphics*[width=\columnwidth]{%
    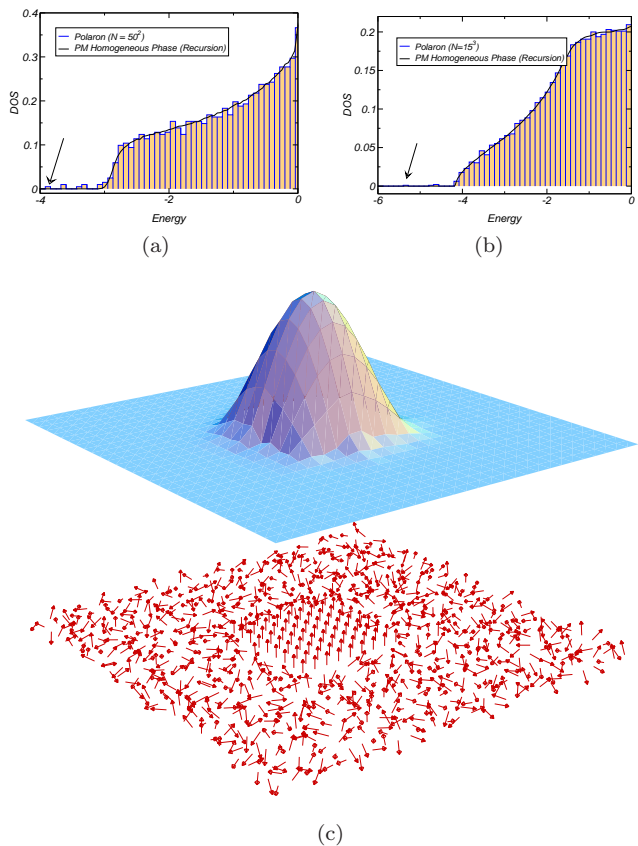}%
    \label{fig:MagPol-DOS-and-Spectrum-3}
  }
  \caption{
    Comparison of the disorder-averaged electronic DOS for the homogeneous
    \ac{PM} phase (line) with the exact spectrum obtained for the same
    configuration  with a polaronic region or size $R$ (histogram). Arrows
    highlight  the lowest  bound state. The results in
    \subref{fig:MagPol-DOS-and-Spectrum-1} are for 2D and $R=9$,
    and in \subref{fig:MagPol-DOS-and-Spectrum-2} for 3D, $R=5$.
    The histogram was obtained from the full numerical diagonalization of the
    polaronic configuration. The continuous line represents the bulk \ac{DOS}
    obtained with the recursion method.
    For this comparison we neglected the Berry phase 
    in the hopping (i.e. we used $a_{ij}\rightarrow\abs{a_{ij}}$).
    Panel \subref{fig:MagPol-DOS-and-Spectrum-3} shows one of those 2D spin
    configurations in the polaronic phase (bottom) together with the
    corresponding exact wavefunction of the lowest bound state (top). 
  }
  \label{fig:MagPol-DOS-and-Spectrum}
\end{figure}
%

There are several reasons to expect it to be a good approximation: 
(i) the electron density is very small, meaning that the overlap of the wave
functions of self-trapped electrons (and thus the polarons) should be
negligible; 
(ii) even if one could devise a full, self-consistent, solution to the problem,
the exact energy of the bound state is expected to differ from the one used in
this approximation only by numerical factors of $\mathcal{O}(1)$. Specific
confirmation for this can be found, for instance, in the 1-D 
calculations of \citet{Pathak:2001}, who find that the two energies (calculated
using the same kind of variational wave function used in the current work and
the exact energy obtained numerically) differ by less than 3~\% in the limit
relevant to our discussion.
;
(iii) numerical calculations strongly favor this approach. To dwell a while on
this last point, we studied the effect of a region of full polarization embedded
in a PM background on the electronic spectrum. Performing exact diagonalizations
of the \ac{DEM} using spin configurations like the one depicted in
Figs.~\ref{fig:MagPol-Polaron-Picture} and \ref{fig:MagPol-DOS-and-Spectrum-3},
one finds, as a result, the appearance of well defined bound states below the
continuum band. 
This is clearly seen in Figs.~\ref{fig:MagPol-DOS-and-Spectrum-1} and
\ref{fig:MagPol-DOS-and-Spectrum-2}, where the bound states appear with energies
and degeneracy coinciding with the value expected for a finite box in
$d\textrm{-dimensions}$:
$-2dt\sum_{\mu=1}^{d}\cos\left(\frac{\pi}{R+1}n_{\mu}\right)$. At the same time,
inspecting particular realizations of disorder as the one in
Fig.~\ref{fig:MagPol-DOS-and-Spectrum-3}, one concludes that the wave function
is clearly localized within the polarized region, thus supporting our trial
function selection for the evaluation of the electronic energy. 

The stability of the magnetic polaron is determined by the condition
$\Delta\mathcal{F}\left(R_{\textrm{eq}},T\right)<0$, $R_{\textrm{eq}}$ being the
value that minimizes the free energy~\eqref{eq:MagPol-DF-WithConfEntropy}.
Important insights can be extracted from the direct analytic results obtainable
in its continuum version (i.e. $R\gg 1$). In this case the equilibrium radius is
\begin{equation}
  R_{\textrm{eq}}\left(T\right) \simeq 
    \left[\frac{2t\pi^{2}}{T\log\left(2S+1\right)}\right]^{1/5}
  \label{eq:MagPol-Req}
  \,,
\end{equation}
increasing at low temperatures as $T^{-1/5}$. This power law behavior of
$R_\text{eq}$ with temperature is reminiscent of the one found in the context
of magnetic semiconductors \cite{Kasuya:1970}.
Using this result in the stability
condition, one finds the stability temperature, $T_{m}$, below which the
polaronic phase appears:
\begin{equation}
  T_m = \frac{8}{25\pi^{3}\log\left(2S+1\right)}\sqrt{\frac{2}{5}}t
  \,.
  \label{eq:MagPol-Tm}
\end{equation}
These two results convey the essential information for the physical situation as
one reduces the temperature of our system. The high-temperature phase is
characterized by \ac{PM} order until $T$ decreases below $T_m$, at which point,
the \emph{``entropic pressure''} is not enough to counteract the energy gain and
the polaronic phase is stabilized. The transition is sharp as a consequence of
the one-particle nature of the free energy~\eqref{eq:MagPol-DF-WithConfEntropy},
and the polarons set in with a finite radius
$R_{\textrm{eq}}\left(T_{m}\right)$. Continuing the decrease in temperature, the
polaron radius increases until the overlapping probability can no longer be
ignored, therefrom arising an instability towards an homogeneous \ac{FM} phase.
An estimate of the Curie temperature, \Tc, can thus be extracted from a
percolation criterion
\begin{equation}
  n_e R_{\textrm{eq}}\left(T_C\right)^{3} \simeq p_{c}
  \label{eq:MagPol-PercolationCriterion}
  \,,  
\end{equation}
yielding
\begin{equation}
  T_C^P\simeq \frac{2 \pi^2 t}{\log(2S+1)} \left(\frac{n_e}{p_c}\right)^{5/3}
  \label{eq:MagPol-TcPercol}
  \,,
\end{equation}
where $p_c$ is the percolation threshold (for the
densities we are interested in $p_c\gg n_e$). For $T_C
< T < T_m$, an anomaly in the paramagnetic susceptibility is expected to signal
the presence of the polarons through an enhanced effective moment.

A natural question can surface at this point regarding the fact that, since
\citet{Gennes:1960}, we know that, using the virtual crystal and the
one-particle approach used above, a transition between uniform \ac{PM} and
\ac{FM} phases should occur. It is therefore of natural interest to investigate
how this magnetic transition is altered by the stabilization of the polaronic
phase. In order to do that we follow the same approach as the one carried by
de~Gennes in evaluating the electronic energy as a function of the
magnetization, obtaining a description in terms of the free energy $\Delta
\Fcal_{FM}(T,M)$. This represents the free-energy difference between the
homogeneous \ac{PM} phase and an homogeneous \ac{FM} phase \cite{Gennes:1960}.

\begin{figure}
  \centering
  \includegraphics*[width=0.90\columnwidth]{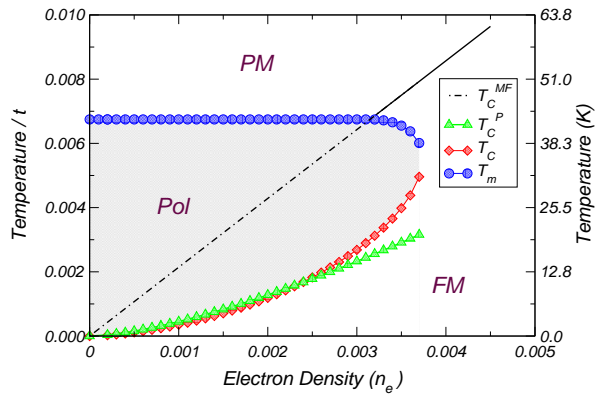}%
  \caption{
    Phase boundaries obtained within the \ac{IPM}. The polaronic phase (Pol)
    appears as a precursor to the ferromagnetism (FM) at low densities (dashed
    region). The dotted-line curve represents $T_C$ obtained in mean field. See
    the text for other notations.
    The left vertical scale is in units of $t$, and the one on the right in
    Kelvin, using $t=0.55$~eV.
  }
  \label{fig:MagPol-PolPhaseDiagSimple}
\end{figure}

Since the \ac{PM} phase is common to $\Delta \Fcal_{FM}(T,M)$ and 
$\Delta \Fcal_\text{Pol}(T,M)$ \eqref{eq:MagPol-DF-WithConfEntropy} we
can calculate the regions of relative stability of the three phases, and draw
the phase diagram shown in Fig.~\ref{fig:MagPol-PolPhaseDiagSimple}.
In this plot, $T_C^{MF}$ represents the line that would be obtained ignoring the
possibility of polaron formation, as de~Gennes did; $T_C^P$ is just the curve
from eq.~\eqref{eq:MagPol-TcPercol} corresponding to the percolation criterion.
Notice that we also included the lines 
$T_C$ and $T_m$ that are the actual transition lines calculated by minimizing
\eqref{eq:MagPol-DF-WithConfEntropy} with respect to the polaron radius and
$\Delta \Fcal_{FM}(T,M)$ with respect to $M$. Even in such simple phase diagram
we observe important physical consequences of the presence of the polaronic
phase, namely: (a) the polaronic phase mediates the transition from an
homogeneous \ac{PM} to an homogeneous \ac{FM} phase, as expected; (b) the
transition temperature calculated in the mean-field approach $T_C^{MF}$ is
notably reduced at low densities by the onset of the polarons; (c) the Curie
temperature obtained with the percolation criterion, $T_{C}^{P}$, gives a very
good estimate of $T_C$ (as follows from the superposition of the two curves in
almost the entire region), supporting the interpretation of the ferromagnetic
transition arising from polaron percolation. 

%
%

As discussed above, in the present framework the
$\textrm{PM}\rightarrow\textrm{Pol}$ transition in this diagram is of first
order and, strictly speaking, the same occurs at the
$\textrm{Pol}\rightarrow\textrm{FM}$ transition, because the phase boundaries
are calculated from the relative stability of the two phases. At first sight
this would mean that, during the $\textrm{Pol}\rightarrow\textrm{FM}$
transition, the magnetization has a discontinuity at $T_{C}$. However, from the
interpretation of the \ac{FM} transition in terms of polaron percolation, one
expects a continuous transition, in the sense that the magnetization of the
system should be weighted by the \emph{mass} of the \emph{infinite percolating
cluster}, which evolves continuously from the percolation threshold. Finally, it
is interesting to notice the order of magnitude of the relevant temperatures and
densities for this phase. If, for definiteness, one assumes that
$t\sim1\,\textrm{eV}$, eqs.~\eqref{eq:MagPol-Tm} and \eqref{eq:MagPol-TcPercol}
reveal that the stability condition is realized for densities of
$\sim 10^{-3}$, and temperatures typically in the dozens of
Kelvin. As is shown in Appendix~\ref{Appendix:MagPol-FiniteBandFilling}, a more
realistic approximation for the electronic energy doesn't appear to modify these
ranges significantly, which somehow restrict the range of materials where the
effect might be realizable.

%
\section{Polaronic Evidence in EuB$_\text{6}$}%
\label{sec:DEM-MagPol-EuB6}

As mentioned earlier, \ac{EuB6} is a magnetic metal with extremely reduced
electron density, and  exhibits all the  characteristic signatures of a
polaronic phase in Raman scattering measurements \cite{Nyhus:1997,Snow:2001}.
Such experiments reveal that the \ac{FM} transition at $T_C \simeq 15$~K is
preceded by an interval of temperatures where the response of the system is
dominated by the presence of magnetic polarons. 

It has been proposed in \Ref~\onlinecite{Pereira:2004} that \ac{EuB6} is very
likely a DEM material in the low density regime, characterized by a
hopping integral of $t=0.55\,\textrm{eV}$, and a carrier density per unit cell
$n_{e}\sim 3\times 10^{-3}$. Subsequent magneto-optical experiments tally with
this \ac{DE}-based interpretation \cite{Caimi:2006}.

If we incorporate such hopping (found to reproduce the
variation in \Wp\ for \ac{EuB6}) in the phase diagram for the \acl{IPM}, the
result is the one shown in Fig.~\ref{fig:MagPol-PolPhaseDiagSimple}, where the
\emph{absolute temperatures} should be read in the \emph{right} vertical axis. 
It is interesting to compare this phase diagram for $n_{e}\simeq0.003$ with the
experimental evidence from the spin-flip Raman scattering results of
\Ref~\onlinecite{Nyhus:1997}.
The experiment reveals a polaronic
region being stabilized essentially in the same temperature range as the one in
Fig.~\ref{fig:MagPol-PolPhaseDiagSimple} for the appropriate densities. This
seems to show that our simple description of the polaronic phase captures the
essential details of the polaron physics in \ac{EuB6}. In particular, we
emphasize that, the calculation of the Curie temperature based on the de~Gennes
approach (the dotted portion of the straight line in
\fref{fig:MagPol-PolPhaseDiagSimple}), clearly overestimates the
actual $T_{C}$ by a factor of 3 or more inside the polaron stability range. As a
matter of fact, placing ourselves at a density $n_e=0.003$ in the diagram of
Fig.~\ref{fig:MagPol-PolPhaseDiagSimple}, we find $T_C \simeq 17$~K. This lies
noticeably close to the experimental \Tc, without adjusting parameters.

%
\section{The Problem of Phase Separation}%
\label{sec:DEM-PhSeparation}

The independent polaron model discussed before, is based on various
approximations that stem from the low electronic density of the systems we aim
to describe. In particular, the same assumption for the electronic energies used
by de~Gennes was employed. One of the implications of approximating
eq.~\eqref{eq:MagPol-Eel_Full} by $E_{el}(M,n_{e})\approx
E_{b}(M)n_{e}$ is that the Curie temperature calculated in mean-field
for homogeneous PM and FM phases is simply proportional to the electronic
density. This happens because the electronic density, $n_e$, appears together
with the only energy scale in the problem, $t$.
                                                                   
But a more serious question regarding the double exchange in the low density
limit is the problem of \acf{PS}. It is know that the \ac{DEM} is unstable
towards phase separation at low carrier densities, even without additional
\ac{AFM} (superexchange) couplings between the lattice spins
\cite{Nagaev:1998,Yunoki:1998,Arovas:1999,Kagan:1999,Alonso:2001a,Alonso:2001b}.
In order to study this aspect of the model in more detail, we will abandon the
previous single particle approximation for the electronic energy,
and calculate this quantity exactly within a \ac{HTA} discussed next.

\subsection{Hybrid Thermodynamic Approach
\label{sec:DEM-Disorder-Thermodynamics}}  

%
\begin{figure}
  \centering
  \includegraphics*[width=0.45\textwidth]{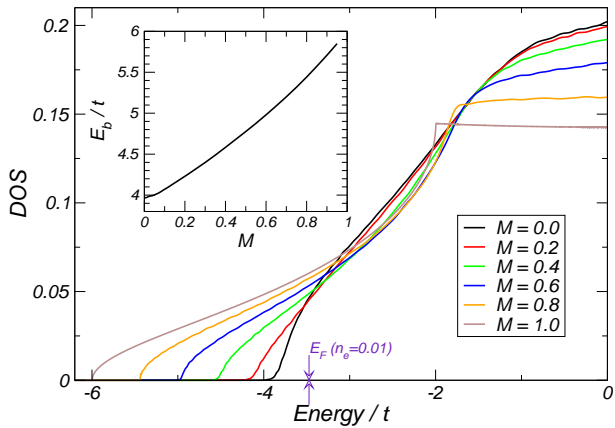}
  \caption{
    Numerical \ac{DOS} for the \ac{DEM} at different magnetizations, as defined
    in eq.~\eqref{eq:DOS-Average-DOS-Notation}. The arrows signal the
    position of $E_F(\mathcal{M}=0)$ for an electron concentration of $n_e =
    0.01$.
    The inset is a plot of the band edge as a function of the local
    magnetization.
  }
  \label{fig:DEM-DOS-DE-Demo}
\end{figure}
%

The \ac{DE} Hamiltonian \eqref{eq:DEM-DE-Hamiltonian-2} represents a
system of classical spins that do not interact directly via a conventional
exchange term. Their interaction comes indirectly from the fact that certain
configurations of the local spins will minimize the energy of the electron gas.
At absolute zero temperature, it is more or less evident that the ground state
corresponds to ferromagnetism for any density of electrons
\cite{Tsunetsugu:1997}. For other than this specific case, thermodynamics comes
into play. All relevant quantities follow from the partition function
\begin{equation}
  \Xi = \int \mathcal{D}\Si \tr 
        \Bigl[\exp \bigl(-\beta (\mathcal H(\Si) - \mu N) \bigr) \Bigr]
  \label{eq:DEM-Exact-Part-Function}
  \,.
\end{equation}
The integral spans all local spins, \Si, and the trace is over the electronic
degrees of freedom, for which $N$ is the number operator. The electrons can be
easily traced out in this grand canonical ensemble yielding an (exact) effective
spin Hamiltonian that reads
\begin{equation}
  \Xi = \int \mathcal{D}\Si \exp\bigl(-\beta H_{eff}(\Si) \bigr)
  \,,
\end{equation}
with
\begin{equation}
  \beta H_{eff} = - \sum_n \log 
    \Bigl[
      1 + \exp \bigl(-\beta (E_n(\Si) - \mu) \bigr)
    \Bigr]
  \label{eq:DEM-Effective-Hamiltonian}
  \,,
\end{equation}
$E_n(\Si)$ being the eigenenergies of the one-electron Hamiltonian. Here,
however, lies the origin of the difficulties that this system poses to
analytical and numerical approaches. The latter are rather notorious, for one
might think that once the effective spin Hamiltonian is written down, the energy
of spin configurations can be calculated and the problem can be tackled with
usual \ac{MC} techniques. That is indeed so formally. Unfortunately, the
effective Hamiltonian --- that is, the energy associated with a given spin
configuration --- requires the knowledge of the electronic eigenstates
associated with such configuration. In other terms, the \ac{MC} methodology
would imply the full re-diagonalization of the electronic problem at every
tentative update of the local spin configuration. This is clearly prohibitive,
imposing an upper limit of $\sim 6^3$ to $8^3$ on the sizes of the systems that
can be thus studied \cite{Calderon:1998,Dagotto_Book:2003}.

Our approach to this problem tries to circumvent these issues through a
compromise in which the electronic problem is solved exactly and the spin
subsystem treated within mean-field \cite{Alonso:2001a}. It hinges upon
the fact that, writing $H_{eff}$ in terms of the total electronic \ac{DOS}
\begin{equation}
  \beta H_{eff} = - \int_{-\infty}^{\infty} \rho(E,\Si) \log 
    \Bigl[
      1 + \exp \bigl(-\beta (E - \mu) \bigr)
    \Bigr] dE
  \label{eq:DEM-Effective-Hamiltonian-DOS}
  \,,
\end{equation}
the dependence on the spin configuration is completely transferred to the
\ac{DOS}. 
Since the treatment of the exact effective Hamiltonian is out of reach, we
resort to the Bogoliubov-Gibbs inequality for the canonical free energy:
\begin{equation}
  \mathcal{F} \equiv -T \log(\Xi) \le \average{H_{eff}}_t - T \mathcal{S}_t
  \label{eq:DEM-Boboliubov-Gibbs}
  \,.
\end{equation}
Here $\average{...}_t$ means the averages are calculated with a \emph{trial}
statistical operator --- other than the canonical Maxwell distribution --- and
$\mathcal{S}_t$ is the associated entropy. Since our system is expected to
exhibit either ferro or paramagnetism, the simplest suitable choice is the one
generated by the uniform mean-field Hamiltonian
\begin{equation}
  \mathcal{H}_t = - h \sum_i^N \Si^z
  \label{eq:DEM-MF_Hamiltonian}
  \,,
\end{equation}
where $h$ is a variational parameter used to minimize the inequality
\eqref{eq:DEM-Boboliubov-Gibbs}. Since all averages are now done with regard to
this $\mathcal{H}_t$, we have:
\begin{align}
  M \equiv \average{\Si}_t & =  
    \int \mathcal{D}\Si \exp(-\beta \mathcal{H}_t) \Si
    \nonumber\\
    & = \left( \coth(\beta h) - \frac{1}{\beta h} \right)\,\vec{u}_z 
    \nonumber\\
    & = \mathcal{L}(\beta h) \,\vec{u}_z 
  \label{eq:DEM-Langevin}
  \,,
\end{align}
where $\mathcal{L}(x)$ is the familiar Langevin function,
\begin{equation}
  - T \mathcal{S}_t = \log\left[\frac{\sinh(h)}{h}\right] - h M
  \label{eq:DEM-MF-Entropy}
  \,,
\end{equation}
and
\begin{equation}
  \beta \average{H_{eff}}_t = - \int_{-\infty}^{\infty} \average{\rho(E,\Si)}_t
\log 
    \Bigl[
      1 + \exp \bigl(-\beta (E - \mu) \bigr)
    \Bigr] dE
  \label{eq:DEM-Effective-Hamiltonian-DOS-MF}
  \,.
\end{equation}
Therefore, the trouble in calculating the equilibrium free energy now boils down
to the computation of $\bigl\langle\rho(E,\Si)\bigr\rangle_t$. We know
\cite{Haydock:1972a,Haydock:1975a,Pereira:2006PhD} that the recursive method can
be used to obtain the exact $\rho(E,\Si)$ for a given configuration $\{\Si\}$.
It is then a matter of straightforward statistics to obtain the averaged
\ac{DOS}. Hence, the electronic problem is still treated exactly for every
configuration of local spins. These configurations are generated with the
probability distribution $\sim \exp(-\beta h \sum_i \Si^z)$ and, since $\beta h$
and $\mathcal{M}$ are univocally related through \eqref{eq:DEM-Langevin}, we
simplify the notation and write
\begin{equation}
  \bigl\langle\rho(E,\Si)\bigr\rangle_t \equiv \rho(E,M)
  \label{eq:DOS-Average-DOS-Notation}
\end{equation}
whenever we refer to the \emph{\ac{DOS} averaged over configurations of disorder
compatible with an average magnetization $M$}.

We also recall that when $E_F \gg T$, the logarithm in
\eref{eq:DEM-Effective-Hamiltonian-DOS-MF} can be replaced by $\Theta(\mu-E)$
meaning that the thermal excitations of the electronic subsystem can be
neglected to a great extent. According to our earlier discussion, we are
interested in cases for which $t\sim 1$~eV and $\Tc \ll t$. This suggests a
``zero-temperature'' description of the electronic system, the thermal/entropic
effects being assigned entirely to the spin subsystem \cite{Endnote-4}.

\subsection{Canonical Free Energy and PhaseDiagram}%
\label{sec:PS-Canonical_Fenergy}

\begin{figure}
  \centering
  \includegraphics*[width=0.90\columnwidth]{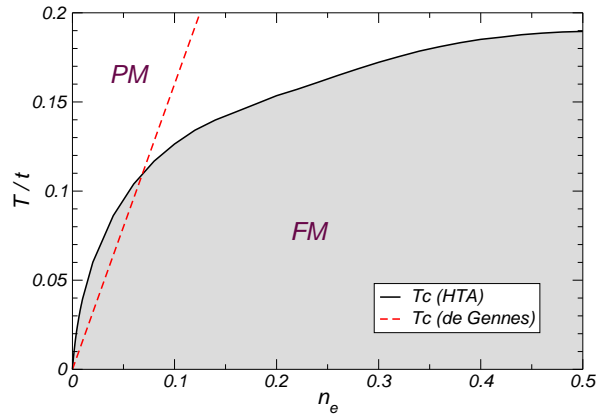}%
  \caption{
    Phase diagram of the DEM obtained within the \ac{HTA}. For comparison, the
    result for $T_{C}$ obtained by de~Gennes is also plotted (dashed line)
    [cfr. \Ref~\onlinecite{Alonso:2001a}].
  }
  \label{fig:PS-PhDiag_DEM_VMF_Large}
\end{figure}

Unless whenever stated otherwise, throughout this section we will be concerned
only with the homogeneous phases (\ac{PM} and \ac{FM}) of the \ac{DEM}. In the
\ac{HTA} one tries to trace the electrons out of the problem and obtain an
effective Hamiltonian for the lattice spins, in such a way that the partition
function becomes simply \eqref{eq:DEM-Effective-Hamiltonian}:
\begin{equation}
  \Xi = \int \mathcal{D}\Si \exp\bigl(-\beta H_{eff}(\Si) \bigr)
  \,. \nonumber 
\end{equation}
This effective Hamiltonian contributes to the total free energy through
eqs.~\eqref{eq:DEM-Boboliubov-Gibbs}, \eqref{eq:DEM-MF-Entropy} and
\eqref{eq:DEM-Effective-Hamiltonian-DOS-MF}, and we have, at the end:
\begin{equation}
  \mathcal{F}(M,n_e) = \int^{E_F(n_e,M)} \epsilon \rho(\epsilon,M) d\epsilon
    - T\mathcal{S}(M)
  \,.
  \label{eq:PS-F-Variational-T0}
\end{equation}
The free energy in eq.~\eqref{eq:PS-F-Variational-T0} should be minimized with
respect to $M$ to obtain the equilibrium free energy $\mathcal{F}_{eq}(n_e)$. We
work at constant electron density and our focus in the cases where $T\ll t$ is
implicitly assumed in the omission of the electronic entropy. The phase diagram
that emerges from the minimization of \eqref{eq:PS-F-Variational-T0} is drawn in
Fig.~\ref{fig:PS-PhDiag_DEM_VMF_Large} and reproduces \cite{Endnote-5}
the results obtained by \Ref~\onlinecite{Alonso:2001a}.
%
%
The PM--FM transition happens to be continuous for all densities and the
deviations from the result obtained with the \emph{bottom of the band},
de~Gennes-like, approximation are quite evident. It turns out, however, that
this phase diagram is incomplete.

\subsection{The Essence of the Problem \label{sec:PS-EssenceProblem}}

Reflecting the relative stability of the two homogeneous phases, the plot in
Fig.~\ref{fig:PS-PhDiag_DEM_VMF_Large} says nothing regarding density
fluctuations. To determine whether the system is unstable in relation to density
fluctuations, we need to look at the behavior of the equilibrium free energy,
$\mathcal F_{eq}(n_e)$, namely its dependence on the electronic density. 
%
\begin{figure}[b]
  \centering
  \includegraphics*[width=0.9\columnwidth]{%
    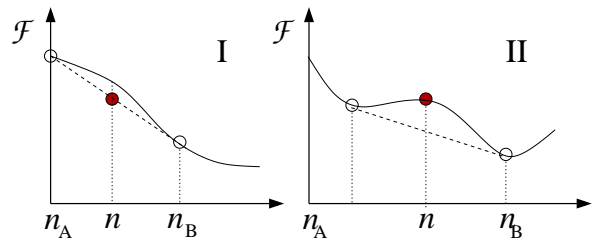}
  \caption{
    Free energy for a system unstable towards density fluctuations. The system
    is unstable for $n_A < n < n_B$ and segregates between regions of density
    $n_A$  and $n_B$. The Maxwell construction restoring the curvature of
    $\mathcal F_{eq}(n_e)$ is represented by the dashed lines.
  } 
  \label{fig:PS-Maxwell_Constr}
\end{figure}
%
Thermodynamic stability requires $\mathcal F_{eq}(n_e)$ to be a globally upward
convex function, so that the compressibility, $\kappa = n^{-1} \partial n
/\partial \mu$, is never negative. Whenever this condition is violated and the
density of the total system is kept constant, it will naturally segregate into
two distinct phases in such a way that guarantees the restoration of convexity
in the resulting free energy. The equilibrium state in this phase-separated
region can be obtained by the so-called Maxwell construction, which,
geometrically, is tantamount to substituting the underlying $\mathcal
F_{eq}(n_e)$ by the envelope of all inferior tangents \cite{Callen:1985}. A
sketch of the situation is presented in Fig.~\ref{fig:PS-Maxwell_Constr}.

\begin{figure}[tb]
  \centering
  \includegraphics*[width=1.0\columnwidth]{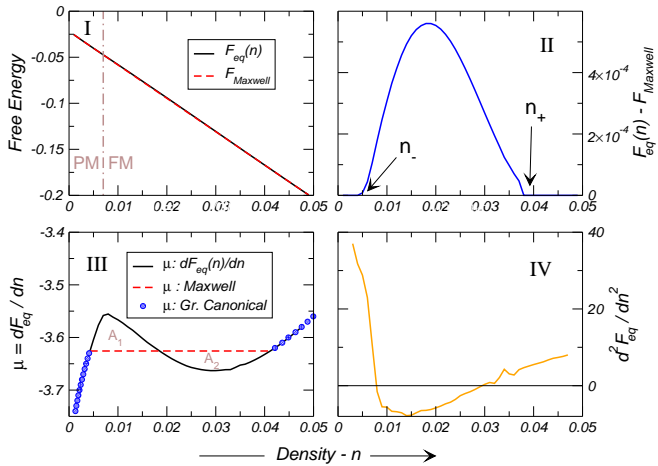}%
  \caption{
    Details of the Maxwell construction for the DEM at $T=0.03t$ . The result of
    the construction is barely discernible from the underlying
    $\mathcal{F}_{eq}(n_e)$ (I) and, for clarity their difference is plotted on
    panel II. For comparison, the chemical potential extracted from the results
    in I is shown together with calculations in the grand canonical ensemble
    (III). Panel IV shows $\partial^{2}\mathcal{F}_{eq}/\partial n_{e}^{2}$ . 
  }
  \label{fig:PS-Max_Constr_Detail}
\end{figure}

It so happens that, when the behavior of the isothermals of $\mathcal
F_{eq}(n_e)$ for the DEM calculated using eq.~\eqref{eq:PS-F-Variational-T0} is
analyzed, the signature of this instability emerges at low densities by the
violation of the global convexity. A typical result is shown in detail in
Fig.~\ref{fig:PS-Max_Constr_Detail}.
%
%
Since there is a considerable amount of information in this figure let us go
through it in detail (everything is calculated at a constant temperature,
$T=0.03t$). 

In the first panel (I) we show the equilibrium free energy $\mathcal
F_{eq}(n_e)$ as a function of the electronic density $n_e$ (continuous/black
line). Despite looking like a straight line, it does have a slight upwards
curvature at the lowest densities, downwards at intermediate densities, and
upwards again at higher densities. This can be seen in the curve
$\partial^{2}\mathcal{F}_{eq}/\partial n_{e}^{2}$ plotted in panel (IV). Thus,
we have an instability and a Maxwell construction has to be done in order to
find the \emph{true} equilibrium free energy of the system. The Maxwell
construction is the dashed/red line in panel (I). Since the effect is rather
subtle, we plot the difference between $\mathcal F_{eq}(n_e)$ and the free
energy after the Maxwell construction in panel (II). In panel (I) we also show
the density corresponding to the PM--FM transition at this temperature, signaled
by the dot-dashed vertical line (cfr. Fig.~\ref{fig:PS-PhDiag_DEM_VMF_Large}).
If we call this particular density $n_e(T_C=0.03)$, then it is clear that $n_{-}
< n_e(T_C=0.03) < n_{+}$. Then, although the relative difference between the
free energies is only $\sim 1\%$, it is qualitatively significant because the
system exibiths coexistence of \ac{PM} and \ac{FM}, each of the coexisting
thermodynamic phases having its own electron density. 

In panel (III) we present the chemical potential calculated in three different
ways. The first one (black/continuous) is simply the curve corresponding to
$\partial F_{eq}(n)/\partial n$, and the instability is also clearly seen here
since $\mu(n)$ should be monotonously increasing. The dashed (red) curve is
the chemical potential resulting from the free energy after the Maxwell
construction. Below $n_{-}$ and above $n_{+}$ the result is the same, but in
between it is simply an horizontal line. Altough not shown in the figure, the
position of this horizontal part is such that the areas $A_1$ and $A_2$ are
exactly equal, as it should happen \cite{Endnote-6}. The third curve (circles)
shows the chemical potential obtained my minimizing
the free energy in the grand canonical ensemble. In this case, since not $n$ but
$\mu$ is kept constant, the phase separation instability arises naturally
through a discontinuity in the $\mu(n)$ curve. The discontinuity appears
precisely in the region of densities where the Maxwell construction is in
effect, and is just another way to observe this phenomenon.  

%
\begin{figure}[b]
  \centering
  \subfigure[][]{
    \includegraphics*[width=0.45\textwidth]{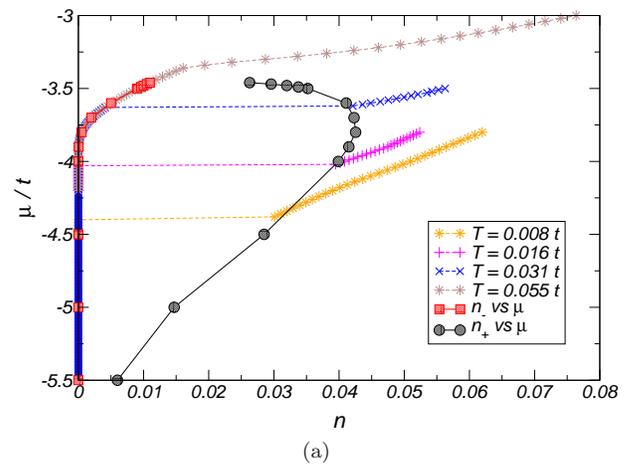}%
    \label{fig:PS-Mu_vs_n_Diagram}
  }
  \subfigure[][]{
    \includegraphics*[width=0.45\textwidth]{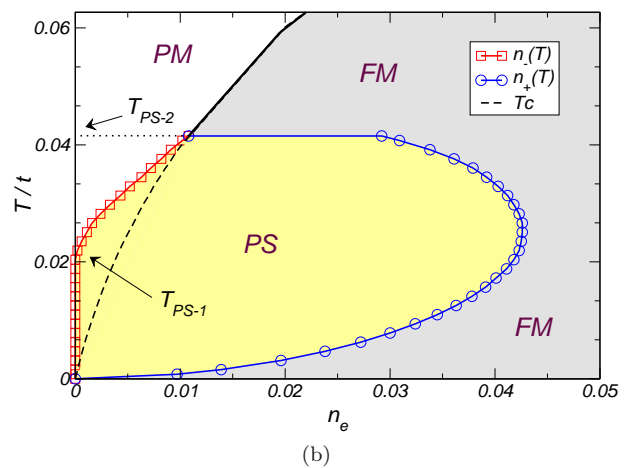}%
    \label{fig:PS-PhDiag_PS_Detail}
  }
  \caption{
    \subref{fig:PS-Mu_vs_n_Diagram} Isothermals in the $\mu-n_{e}$ plane for
    selected temperatures. The coexistence curves $n_{-}$ (squares) and $n_{+}$
    (circles) were obtained from the Maxwell construction, whereas the
    $\mu(n_{e})$  curves are calculated in the grand canonical ensemble, for
    comparison.
    \subref{fig:PS-PhDiag_PS_Detail} Detail of the low density region of the
    updated $T-n_e$ phase diagram. The phase separated region (PS) is bounded by
    the  curves $n_{-}(T)$ and $n_{+}(T)$ obtained from the Maxwell
    construction. 
  }
  \label{fig:PS-PhaseDiagram-with-PS}
\end{figure}
%

Actually it is quite instructive to pursue in some more detail this
complementarity between the canonical and grand canonical treatments for this
particular case. Our scenario is completely analogous to the well known behavior
of the van~der~Waals
isothermals for the liquid--gas transition \cite{Callen:1985}. 
This is best understood with reference to the plots of $\mu(n_{e})$ in
Fig.~\ref{fig:PS-Mu_vs_n_Diagram} at several temperatures. Just as in the $P-v$
phase diagram of the gas, the $\mu(n_{e})$ curve is monotonous for temperatures
above some  critical value. Below this point, the compressibility of the system
($dn_{e}/d\mu$) becomes negative in some density interval implying an
instability. The coexistence curve 
is then defined by $n_{-}$ and $n_{+}$ obtained from the Maxwell construction
and, naturally, coincides with the jump in the density obtained in the grand
canonical calculation. Despite the analogy and resemblance of the coexistence
curve for the DEM and the van der Waals gas, there is an important qualitative
detail not present in the latter
case: the coexistence curve for the \ac{DEM} is re-entrant.

The stability analysis was performed for all the temperatures and densities
shown in the phase diagram of Fig.~\ref{fig:PS-PhDiag_DEM_VMF_Large}. The system
is unstable towards phase separation at low densities, and the qualitative
behavior of the thermodynamic functions is always the one discussed above. As a
consequence, one obtains the updated phase diagram presented in
Fig.~\ref{fig:PS-PhDiag_PS_Detail}.

\begin{figure}
  \centering
  \includegraphics*[width=0.8\columnwidth]{%
      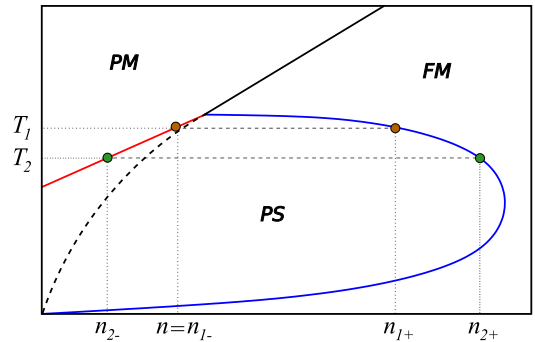}%
  \caption{
    Schematic representation of Fig.~\ref{fig:PS-PhDiag_PS_Detail}. 
  }
  \label{fig:PS-PS_Details}
\end{figure}

The information conveyed by this diagram can be translated as follows: for
densities above $n_{e}\sim0.04$, the system is homogeneous and exhibits a
magnetic transition at $T_C(n_e)$. If the density is lower than $n_e\sim0.01$
the system is homogeneous and \ac{PM} at high temperatures until $n_{-}(T)$ is
reached. At that point the homogeneous phase is no longer sustainable and two
phases start to coexist: one with density $n_{-}(T)$ and \ac{PM} together with
another of density $n_{+}(T)$ and \ac{FM}. Below a temperature signaled as
$T_{\textrm{PS-1}}$, $n_{-}(T)=0$ and thus the \ac{PM} portion of the system is
devoid of electrons \cite{Endnote-7}. If $0.01\lesssim n_{e}\lesssim0.04$, the
re-entrant nature of the coexistence
curve means that the system can become an homogeneous ferromagnet below
$T_{C}(n_{e})$ but still segregate at lower temperatures whenever the condition
$n_{e}=n_{+}(T)$ is met. In any case, the magnetization of the system is always
continuous~--~a simple consequence of the relation between the densities and
volume fractions of each phase:
\begin{equation}
  n_e = n_{-} \frac{V_{-}}{V} + n_{+}\frac{V_{+}}{V}
  \,.
  \label{eq:PS-Volume_Fractions}
\end{equation}
This constraint introduces some peculiarities regarding the nature of the phase
separated state. To understand that, in Fig.~\ref{fig:PS-PS_Details} we draw a
sketch of the phase separated region in the phase diagram of
Fig.~\ref{fig:PS-PhDiag_PS_Detail}.
%
%
Using this sketch as reference, assume that our system is initially at some high
temperature $T>T_1$ and has a given density $n$. Under these circumstances, the
phase diagram states that the equilibrium corresponds to an homogeneous \ac{PM}
phase. We can lower the temperature until $T_1$ is reached, at which point an
instability arises. Exactly at $T=T_1$ there will be a segregation between a
\ac{PM} phase with density $n_{\textrm{1-}}=n$ and another, \ac{FM}, with
density $n_{1+}$. In order to satisfy the constraint
\eqref{eq:PS-Volume_Fractions}, the volume fraction of the \ac{FM} phase will be
$V_{\textrm{1+}}=0$, at $T=T_{1}$. A slight decrease in the temperature from
$T_{1}$ to $T_{2}$ will reorganize the system so that the \ac{PM} phase with
density $n_{\textrm{2-}}$ now coexists with a \ac{FM} phase of density
$n_{\textrm{2+}}$. For $T_{2}$ very close to $T_{1}$, the density of the \ac{PM}
phase is just slightly different from the global density:
\begin{equation}
  n_{\textrm{2--}} = n-\varepsilon
  \,,
\end{equation}
where $\varepsilon$ is a small quantity. This determines the volume fractions to
be
\begin{equation} 
  V_{\textrm{2+}} \approx  \frac{\varepsilon}{n_{\textrm{2+}}-n}
  \,.
  \label{eq:PS-V2+}
\end{equation}

Evidently, this means that most of the electrons still remain in the \ac{PM}
phase, and just a few populate the \ac{FM} regions. But this poses a problem.
The Coulomb interaction will certainly prevent the accumulation of charges in a
very small volume and the system must remain neutral.

\subsection{Electrostatic Suppression of Phase Separation
\label{sec:PS-CoulombSuppression}}

The Maxwell construction is inexpressive with regards to the spatial
organization of the phase separated state. This follows from the fact that the
Maxwell construction for two coexisting phases of densities $n_+$ and $n_-$
amounts formally to saying that,
\begin{widetext}
\begin{equation}
  \mathcal F_\text{Maxwell}(n) = \mathcal F(n_+) \frac{V+}{V} + 
      \mathcal F(n_-) \frac{V-}{V}
    = \mathcal F(n_+) x + \mathcal F(n_-) (1-x)
  \,,
  \label{eq:PS-MaxwellConstruction}
\end{equation}
\end{widetext}
where $x=V_+/V$ and $V_-/V$ represents the volume fraction of the two coexisting
phases. This is simply a linear interpolation between the two values $F(n_+)$
and $F(n_-)$, as illustrated in Fig.~\ref{fig:PS-Maxwell_Constr}. Therefore, the
resulting free energy, $\mathcal F_\text{Maxwell}$, corresponds to the situation
where we have a system made of two independent thermodynamic components. In
particular, the Maxwell prescription above says nothing about the way the system
reorganizes when it phase-separates. This can only come from additional
interaction terms that should be added to the right hand side of
eq.~\eqref{eq:PS-MaxwellConstruction}, in order to include interaction, surface,
boundary and perhaps other relevant effects or correlations. In our case the two
phases have different
electronic densities, both different from the homogeneous density, the latter
satisfying
\begin{equation}
  n = n_+ x + n_- (1-x)
  \,.
\end{equation}
Obviously, with mobile negative charges, the system can indeed adjust itself by
accumulating electrons in some regions, and depleting them from others. But
since the background of positive atomic charges stays essentially immutable and
homogeneous, this means that the \emph{total} charge of each phase is not zero.
Coulomb interactions are therefore crucial.  Under this assumption we now
investigate two important corrections for the free energy in the phase separated
regime.

\subsubsection{Electrostatic Correction
\label{sec:MagPolPS-ElectrostaticTerm}}

\begin{figure}
  \centering
  \includegraphics*[width=0.9\columnwidth]{%
    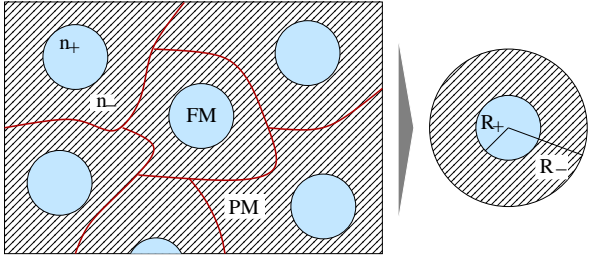}%
  \caption{
    Depiction of the Wigner-Seitz construction discussed in the text.
    Each cell is replaced by an effective spherical cell characterized by the
    same densities.
  } 
  \label{fig:PS-WignerSeitz} 
\end{figure}

Given that we are not assuming any sort of anisotropy, the electrostatic
constraint will most certainly favor the development of bubbles of the \ac{FM},
high density phase, dispersed in the \ac{PM}, low density phase. Such scenario
is schematically depicted in Fig. \ref{fig:PS-WignerSeitz}. 
%
%
The charge density is assumed uniform and continuous inside each \ac{FM} bubble
and across the \ac{PM} background. To calculate the electrostatic energy
associated with this charge distribution we take notice to the fact that, on the
grounds of overall charge neutrality, it should be possible to find an
appropriate neighborhood around each \ac{FM} bubble such that the total charge
on the bubble plus \ac{PM} neighborhood adds to zero. In
Fig.~\ref{fig:PS-WignerSeitz} this is represented by the wavy lines that, in
this way, define cells of charge neutrality. Following Wigner
\cite{Pines:1963}, these cells are replaced by the equivalent \ac{WS} spherical
cell containing the same volume fractions of \ac{FM} and \ac{PM} phases (as
shown on the right-hand side of the diagram), which means that
\begin{equation}
  R_+^3 = x R_-^3
\end{equation}

For each \ac{WS} cell, the total electrostatic energy is calculated considering
three terms
\begin{equation}
  E_C = U_{++} + U_{--} + U{+-}
  \,,
  \label{eq:PS-EC}
\end{equation}
with the first accounting for the electrostatic self-energy of the ``+'' region,
the second the self-energy of the ``-'' region, and the last the mutual
electrostatic interaction between the two. All of them are calculated within
classical electrostatic theory assuming uniform charge distributions in the two
regions. Hence
\begin{equation}
  U_{++} = \frac{3}{5} \frac{Q_+^2}{R_+}
  \,,
\end{equation}
and represents the electrostatic energy of the inner sphere containing the ``+''
region;
\begin{equation}
  U_{--} = \frac{3 Q_-^2}{R_-^3 - R_+^3} 
    \left(
      \frac{R_-^5 - R_+^5}{5} - R_+^3 \frac{R_-^2 - R_+^2}{2}
    \right)
\end{equation}
stands for the energy of the outer shell of the ``-'' region, and
\begin{equation}
  U_{+-} = \frac{3}{2} Q_+ Q_- \frac{R_-^2 - R_+^2}{R_-^3 - R_+^3}
\end{equation}
is the electrostatic interaction between the inner sphere and outer shell. In
the above $Q_\pm$ are the total charges (positive background + electrons) inside
the two regions, and $R_\pm $ the respective radii, as depicted in
Fig.~\ref{fig:PS-WignerSeitz}. For reasons regarding numerical stability, in the
following we will take always $n_- = 0$. Looking at
Fig.~\ref{fig:PS-PhDiag_PS_Detail}, this comes as a natural approximation
because $n_+ \gg n_-$ for $T_{PS1} < T < T_{PS2}$, and becomes exact for $T <
T_{PS1}$. Hence, using the identities
\begin{align}
  Q_+ &= e (n - n_+) V_+ = e n (x-1) V \nonumber\\  
  V_+ &= \frac{4\pi}{3} R_+^3 \nonumber \\
  Q_- &= e (n - n_-) V_- = e n (1-x) V \nonumber\\
  V_- &= \frac{4\pi}{3} (R_-^3 - R_+^3)
  \,
\end{align}
the Coulomb term can be cast as
\begin{equation}
  E_C = \frac{8}{15} e^2 \pi^2 n^2 R_+^5 \frac{2 - 3x^{1/3} + x}{x^2}
  \label{eq:PS-CoulombTerm}
  \,,
\end{equation}
which yields an energy per unit of volume of
\begin{equation}
  \epsilon_C = \frac{E_C}{V} = \frac{2}{5}\,e^2\,n^2\,\pi \,R^2\,
    \frac{2 - 3\,x^{1/3} + x}{x}
  \,.
  \label{eq:PS-CoulombTerm-2}
\end{equation}
For simplicity we replaced $R_+$ by $R$ and keep this lighter notation below.
The bubble radius, $R$, is a variational parameter. The result
\eqref{eq:PS-CoulombTerm-2} implies that, for a given volume fraction of the two
phases, say $x$, the electrostatically favorable situation is to shrink the
\ac{FM} bubbles to an arbitrarily small size, meaning that $R\to0$
\cite{Endnote-8}. 
But this treatment is yet incomplete, inasmuch as the consideration of
finite-sized bubbles of electrons requires another correction, of different
nature, to the ground-state energy.

\subsubsection{Phase Space Correction \label{sec:MagPolPS-PhaseSpaceTerm}}

The electronic contribution to the free energy \eqref{eq:PS-F-Variational-T0} is
calculated in the thermodynamic limit. In the phase separated regime, the
electron rich bubbles are expected to be of relatively small size. Therefore,
one cannot rely on the electronic energy calculated in the thermodynamic limit
and the need to introduce finite size corrections arises. 

The leading correction to the energy of an electron gas confined to a finite
sized volume comes through the correction to the electronic \ac{DOS} of the free
electron gas which, as discussed in
Appendix~\ref{Appendix:MagPol-FiniteSizeCorrection}, leads to a correction to
the ground state energy per electron reading
\begin{equation}
  \frac{E(R, n_+)}{N_e} = \frac{E(\infty,n_+)}{N_e} 
    \left[
      1 + \frac{15}{16} \left(\frac{\pi}{6 n_+}\right)^{1/3} \frac{1}{R}
    \right]  
  \label{eq:PS-PhaseSpaceTerm}
  \,.
\end{equation}
The term in $R^{-1}$ is the correction for each bubble. To obtain the total
correction we just multiply by the number of \ac{WS} cells, obtaining the total
energy per unit volume
\begin{equation}
  \frac{E(R,n_+,x)}{V} = \frac{E(\infty,n_+)}{V} 
    \left[
      1 + \frac{15}{16} \left(\frac{\pi}{6 n}\right)^{1/3} \frac{x^{1/3}}{R}
    \right]  
    x 
  \label{eq:PS-PhaseSpaceTerm-2}
  \,.
\end{equation}
As expected, the phase space correction acts to the effect of rising the ground
state energy of the electron gas. This induces a tendency opposed to the one
embodied in the electrostatic term \eqref{eq:PS-CoulombTerm-2}, and the two
should balance at some optimum value of $R$.

%
\begin{figure}[t]
  \centering
  \includegraphics*[width=0.9\columnwidth]{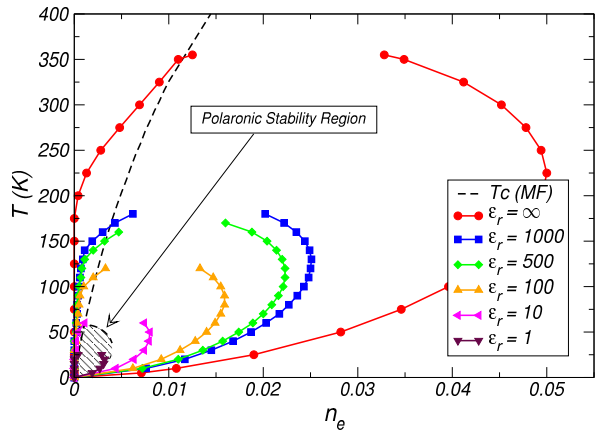}%
  \caption{
    Phase diagram calculated within the \ac{HTA} with a corrected Maxwell
    construction according to eq.~\eqref{eq:PS-FreeEnergy-Corrected}. The arrow
    marks the polaronic stability region, represented by the dashed area at low
    density and temperature. Notice how the region of phase separation lies
    inside the polaronic bubble for dielectric constants of $\varepsilon_r
    \sim 1$. 
  }
  \label{fig:MagPol-PS-Suppression}
\end{figure}
%

The free energy in the phase separated regime \eqref{eq:PS-MaxwellConstruction}
needs to be updated for these two corrections that go beyond the Maxwell
construction. The result is 
\begin{widetext}
\begin{equation}
  \mathcal F_{Maxwell}(n) = 
    \mathcal F(n_+) x + 
    \mathcal F(n_-) (1-x) +
    \frac{2}{5}\,e^2\,n^2\,\pi \,R^2\, \frac{2 - 3\,x^{1/3} + x}{x} +
    \frac{E_{el}(n_+)}{V} \frac{15}{16} \left(\frac{\pi}{6 n}\right)^{1/3}
    \frac{x^{4/3}}{R}
  \label{eq:PS-FreeEnergy-Corrected}
  \,.
\end{equation}
\end{widetext}%
Since the \ac{FM} radius is entering only in the correcting terms, the
minimization with respect to $R$ can be performed at once, and the final result
is
\begin{widetext}
\begin{equation}
  \mathcal F_{Maxwell}(n) =  
    \mathcal F(n_+) x + 
    \mathcal F(n_-) (1-x) + 
    \frac{3}{2^{2/3}} 
      \left(
        \frac{2}{5} \frac{e^2}{a} \pi n_e^2
      \right)^{1/3}
      \left[
        \frac{3}{5}\,t\, n_e (6\pi^2 n_e)^{2/3} \frac{15}{16} 
        \left(
          \frac{\pi}{6 n_e} 
        \right)^{1/3}
      \right]^{2/3}
      \left[
        \frac{2+x-3x^{1/3}}{x^{5/3}}
      \right]^{1/3}
  \label{eq:PS-FreeEnergy-Corrected-Final}
  \,.
\end{equation}
\end{widetext}
This last expression is the free energy per site of the original lattice when
phase separation is in effect. $a$ is the lattice parameter, $n_e$ the electron
density per unit cell of the crystal, and $t$ is the hopping integral.
Naturally, when there is no \ac{PS} instability, $x=1$ by definition and the
above reduces to $\mathcal F(n_e)$ as one certainly expects. In the \ac{PS}
regime, the equilibrium radius of the electron rich \ac{FM} bubbles satisfies
\begin{equation}
  \frac{4}{3}\,\pi\, n_+ R^3 =
    \dfrac{15}{16} \frac{at}{e^2} 
      \left(
        \frac{6 \pi^5 n_e}{x}
      \right)^{1/3}
    \frac{1}{2+x-3x^{1/3}} 
  \,, 
  \label{eq:PS-EquilibriumRadius}  
\end{equation}
and this relation can be used, for instance, to inspect the typical number of
electrons inside each \ac{FM} bubble.

\subsection{Consequences for the Phase Diagram
\label{sec:MagPolPS-Consequences}}

The natural question is now: what happens when the equilibrium free energy is
recalculated with these corrections? Namely, we want to know wheather the
\ac{PS} instability persists when the Maxwell construction is updated according
to \eqref{eq:PS-FreeEnergy-Corrected-Final}. It is useful to have a tuning
parameter that interpolates between the case in
\eqref{eq:PS-FreeEnergy-Corrected-Final} and the previous calculation where
electrostatic and localization effects were disregarded. With that purpose, we
introduce the
dielectric constant, $\varepsilon_r$, that renormalizes the electron charge as
$e^2 \to e^2/\varepsilon_r$ in the expressions above. By varying $\varepsilon_r$
between $1$ and $1000$ the curves in Fig.~\ref{fig:MagPol-PS-Suppression} were
obtained.
In this plot, we are focusing on the low density region of the phase diagram
where \ac{PS} occurs. The red (circles) curve pertains to the case
$\varepsilon_r\to\infty$ (or, equivalently, zero electronic charge) and is again
the result shown before in the phase diagram of
Fig.~\ref{fig:PS-PhDiag_PS_Detail} \cite{Endnote-9}. 
The figure is transparent as to what happens when the Coulomb interaction is
\emph{turned on}: the \ac{PS} region is progressively reduced! Not only that but
it is clear that an overwhelming shrinking of the \ac{PS} region takes place
when $\varepsilon = 1$, which is a reasonable value, considering the low
electronic densities. 

Thus, the consideration of the free energy
\eqref{eq:PS-FreeEnergy-Corrected-Final},  corrected for the effects arising as
a consequence of charge accumulation, leads to the suppression of the \ac{PS}
instability. The electrostatic payoff involved in the segregation leads the
system to phase-separate only at much lower densities and/or temperatures. Just
how low these are is controlled by the effective electronic charge.

\subsection{Phase Separation and Magnetic Polarons
\label{sec:MagPolPS-PS-and-MagPol}}

There is a question of relevance that we have been postponing since the
beginning of this section on the problem of \ac{PS}. In
\sref{sec:DEM-MagPol-EuB6} calculated the phase diagram of
Fig.~\ref{fig:MagPol-PolPhaseDiagSimple} describing the stability conditions
for free magnetic polarons in the \ac{DEM}. From the polaronic phase diagram
follows that magnetic polarons are only stable at considerably low temperatures
and densities. In fact much lower than the temperatures and densities at which
the \ac{PS} instability sets in. The reader might have noticed since
Fig.~\ref{fig:PS-PS_Details} that the density and temperature scales for the
\ac{PS} bubble are much higher than the scales for the polaronic bubble. More
precisely, the polaronic phase lies well inside the \ac{PS} region when the
corrections to the Maxwell construction are ignored. These two regions can be
seen in perspective in Fig.~\ref{fig:MagPol-PS-Suppression}, where the polaronic
stability region is highlighted by the dashed region in the lower left corner,
and completely inside the \ac{PS} region for $\varepsilon\to\infty$.

This is clearly a problem to our arguments concerning the polaronic phase. The
polaronic stability has been determined by studying its relative stability with
regards to an \emph{homogeneous} \ac{FM} phase. The diagram above is saying
that, at such low densities, there are no homogeneous phases --- the system
phase separates! So the study of polaronic stability has a problem. 

But this is only if the electrostatic effects are ignored.  
%
\begin{figure}[tb]
  \centering
  \includegraphics*[width=0.9\columnwidth]{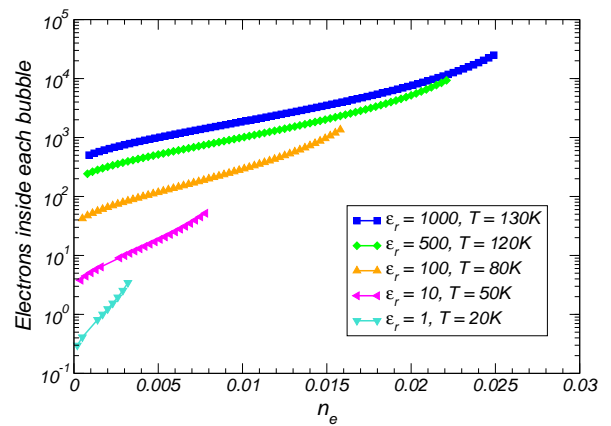}%
  \caption{
    The number of carriers inside each \ac{FM} bubble calculated according to
    \eqref{eq:PS-EquilibriumRadius} for selected temperatures and dielectric
    constants.
  }
  \label{fig:MagPol-R3nFactor}
\end{figure}
%
With their inclusion, the \ac{PS} region retreats to lower and lower densities
and, as the figure documents, for $\varepsilon_r \sim 1$, it rests already
completely inside the polaronic region. So, it seems that the problems above
with the polaronic phase have just diminished. 

The connection between \ac{PS} and magnetic polarons is indeed remarkably close.
The ferromagnetic droplets, associated with a localization energy for the
electrons in a restricted volume, are nothing more than our description of the
magnetic polaron. So, in this sense, the \ac{PS} regime studied here and the
magnetic polarons are different perspectives of the same physical concept. One
of the differences is that, while the magnetic polaron is defined as a \ac{FM}
droplet with a single electron, the \ac{PS} regime allows for droplets with many
electrons. 

To explore this further, it is interesting to know the number of electrons
inside each \ac{FM} bubble in the \ac{PS} regime. Some typical results are
plotted in Fig.~\ref{fig:MagPol-R3nFactor} for the same $\varepsilon_r$ used
before. The most remarkable fact about these curves is that the one pertaining
to $\varepsilon_r=1$ is of the order of unity. Therefore, there is essentially
one electron per \ac{FM} droplet. In addition, the temperatures at which \ac{PS}
occurs for this value of $\varepsilon_r$ are so low that the \ac{FM} droplets
are very nearly full polarization. But, a fully polarized \ac{FM} droplet with
one electron inside is just our definition of magnetic polaron! Thus, the
peculiarities of the phase segregation in this system are completely consistent
with the magnetic polaron picture and, with that respect, the similarity between
the shape of the two phase diagrams (Figs.~\ref{fig:PS-PhDiag_PS_Detail} and
\ref{fig:PS-PhDiag-PolaronFinite-1}) seems hardly coincidental.

\subsection{General Argument Regarding Phase Separation
\label{sec:MagPolPS-GeneralArgument}}

Although we have been focusing so far on the specificities of the phase
separation instability in the \ac{DEM}, phase separation is quite a general
phenomenon in thermodynamics. To conclude our discussion we put forward an
argument showing that, if the contribution of the entropy of the
electronic gas can be neglected in the total free energy, then \emph{phase
separation is ubiquitous in electronic systems whose bandwidth is magnetization
dependent}.
To see how this comes about, recall that the free energy
\eqref{eq:PS-F-Variational-T0} is given by only two contributions: the
electronic ground state energy plus an entropic term attributed uniquely to the
local moments. For reasons that will be clear in a moment, this argument unfolds
more clearly if we work in the grand canonical ensemble, where the electronic
chemical potential, $\mu$, is held constant. In this case the free energy
\eqref{eq:PS-F-Variational-T0} is simply replaced by the grand canonical
potential, and is akin to replacing $E_F\rightarrow\mu$ and
$\varepsilon\rightarrow\varepsilon-\mu$ there. 

Consider now the following facts, regarding the relevant thermodynamic
quantities seen as functions of the magnetization:
\begin{enumerate}
  \item The magnetic entropy \eqref{eq:DEM-MF-Entropy} is monotonous and
downward convex, for all the domain of $M$; 
  \item For a given chemical potential, $\mu$, the electronic energy is
monotonous and downward convex throughout the entire domain of variation of $M$;
  \item The electronic density is monotonous and upward convex. 
\end{enumerate}
Statement 1 follows from the fact that $\mathcal S(M)$ is a proper thermodynamic
entropy for a magnetic system, having all the required analytical properties. 
%
\begin{figure}
  \centering
  \includegraphics*[width=0.7\columnwidth]{%
    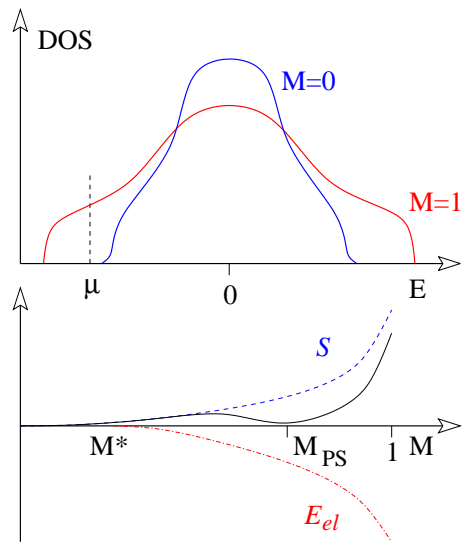}%
  \caption{
    Schematic variation of \ac{DOS}, entropy and electronic energy with $M$ at
    constant $\mu$. In the top panel we illustrate the changes in the \ac{DOS}
    with increasing magnetization, and the corresponding increase in bandwidth.
    In the lower panel we illustrate the variation of the electronic energy 
    and magnetic entropy, and how they combine to yield  the free energy
depicted
    by the black (solid) line.
  }
  \label{fig:Appendix-PSargumentDemo}
\end{figure}
%
Point 3 is a trivial consequence of the electronic density being the integrated
\ac{DOS}.
Point 2 can be understood from the fact that the electronic bandwidth is
monotonous with $M$ (cfr. \fref{fig:DEM-DOS-DE-Demo}). There is a subtlety
however in that, if $\mu$ happens to be below the band edge at $M=0$, then, the
electronic energy will be identically zero until some critical magnetization,
say $M^*$, is reached for which the band edge coincides with $\mu$. For higher
magnetizations, the energy decreases and the overall shape is as depicted in
Fig.~\ref{fig:Appendix-PSargumentDemo}. Since the same plateau is present in the
electronic density, for exactly the same reasons, the result for the grand
canonical potential will be something like the solid curve drawn schematically
in the bottom frame panel of Fig.~\ref{fig:Appendix-PSargumentDemo}. Evidently,
there will be a temperature at which the minimum of this curve exactly touches
the horizontal axis (as depicted), thus precipitating a first order transition.
At the precise temperature, $T$, at which this happens, the system stays
undecided as to which state it should have because the thermodynamic potentials
at $M=0$ and $M=M_{PS}$ are degenerate. Since $\mu$ is kept constant, $M=0$ and
$M=M_{PS}$ correspond to different electronic densities. This is but our phase
separation instability seen from the grand canonical perspective.

The important thing here is that nothing in this argument mentions the details
of the specific model under consideration, and therefore is valid as long as the
basic assumptions remain valid. In particular, the magnetization is as good as
any other suitable thermodynamic parameter, and, thus, the arguments extends to
any appropriate classical variable coupled to the electronic energy as the
magnetization is in our specific case. 

That phase separation has to emerge always follows from the fact that one can
always place $\mu$ in between the band edges at $M=0$ and $M=1$. So, if
temperatures are so low that the fermionic entropy can be disregarded, this kind
of treatment should always yield a phase separated regime at low densities.

%
\section{The regime of ultra low densities}%
\label{sec:WignerCrystal}

Having clarified the issue of phase separation and its connections with the
polaronic phase, we return now to the polaronic description. The consideration
of electron-electron interaction, as done in the previous section, amounts to
effectively describing the electrons in terms of the Hamiltonian 
\begin{align}
  \mathcal{H}_{DE} &= -\sum_{\left\langle i,j\right\rangle}
  a_{ij}d_{i}^{\dag}d_{j}+\textrm{h. c.} \nonumber\\
  &+ e^2 \sum_{i>j}
  \frac{(n_{i}-n_{e})(n_{j}-n_{e})}{|{\bf r}_i-{\bf r}_j|}
  \,,   
  \label{eq:H_DE_Coulomb}
\end{align}
where the presence of the Coulomb term is now explicit. The presence of the full
Coulomb interaction in \eref{eq:H_DE_Coulomb} is to be understood in
connection with the case of extremely reduced electron density. In a
conventional Fermi liquid the high density of electrons makes the screening
process very effective, and the effect of the electron-electron interactions
can be absorbed into the renormalization of physical quantities such as
the effective mass. In a very diluted electron gas the Coulomb
interaction cannot be addressed meaningfully in this way. In the language
of the one component plasma, this can be understood with reference
to the dimensionless parameter $r_{s} = r_o/a_0$, where
$r_o = (3/4\pi\rho)^{1/3}$ ($\rho$ being the volumetric density of electrons) is
the average distance between electrons, and $a_0$ is the Bohr radius. While 
the kinetic energy scales as $1/r_{s}^{2}$, the potential energy varies as
$1/r_{s}$, and, therefore dominates in the low density ($r_{s}\to\infty$)
regime \cite{Pines:1963}. 

It is well known since Wigner \cite{Wigner:1934} that, under such circumstances,
the electrons arrange themselves in a regular lattice, and electron
localization occurs.
If a so-called Wigner crystal is realizable in a system described
by \eref{eq:H_DE_Coulomb} then, as in any solid, there is
zero point motion of the electrons about their equilibrium
positions. If the electron itinerates among several unit cells during
this zero point motion, the magnetic coupling, $J_{H}$, leads to the local
polarization of the lattice spins, producing a \emph{bound}
magnetic polaron and generating a polaronic Wigner crystal. 
This process is obviously limited by the melting of the electronic solid and,
thus, the question arises of how to describe these two tendencies for the
polaron formation and the interplay of Coulomb and magnetic interactions. We
present estimates regarding the stability of this polaronic Wigner crystal
below.

%
\begin{figure}
  \centering
  \includegraphics[width=0.9\columnwidth]{%
    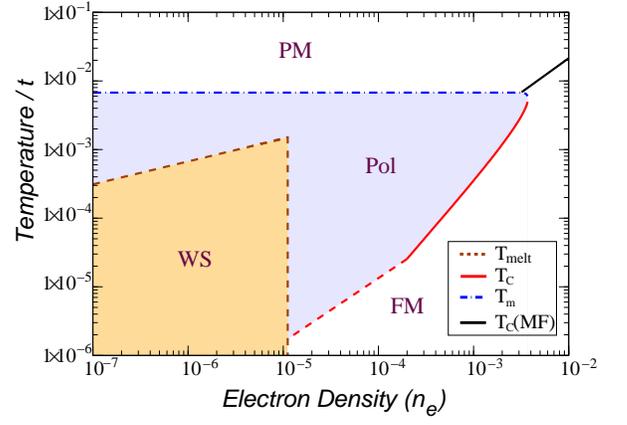}
  \caption{
    Phase diagram of the diluted DEM model as a function
    of temperature and density ($t=0.5\,\textrm{eV},a=4\,\textrm{\AA}$):
    paramagnetic (PM), polaronic (Pol),  
    polaronic Wigner solid (WS), and ferromagnetic (FM).
    Notice that the axes are presented on a logarithmic scale
    (cfr.~\fref{fig:MagPol-PolPhaseDiagSimple}).
  }
  \label{fig:PhDiag_WS_Polarons} 
\end{figure}
%

\subsection{Polaronic Wigner Crystal}
\label{sec:PolaronicWignerCrystal}

Following the Wigner-Seitz approximation \cite{Pines:1963}, the 
Wigner crystal unit cell (much larger than the original
lattice spacing, $a$) is approximated by an electrically neutral
spherical volume, inside which the ionic charge density is homogeneous. 
The electrostatic potential energy then
depends only upon $r$: the distance of the electron from the center
of the cell. The Hamiltonian for an electron in this
uniform charge and spin background is then:
\begin{equation}
  \mathcal{H}_{W} = -6 t - \frac{3}{2}\frac{e^{2}}{r_{o}} 
  + \frac{p^{2}}{2m}+\frac{1}{2}m(\omega_{e}^{2} + \omega^2) r^{2}
  \,,
  \label{eq:H_WignerCrystal}
\end{equation}
where $p$ is the electron momentum,  $m=1/(2 a^{2} t)$ the 
effective electron mass, and $\omega_{e}^{2}=\omega_{p}^{2}/3=e^{2}/mr_{o}^{3}$,
where $\omega_p$ is the plasma frequency. In \eref{eq:H_WignerCrystal}
$\omega$ is the frequency of the confining potential due to the DEM mechanism,
and is a variational parameter. The radius of the magnetic polaron in the Wigner
crystal relates to $\omega$ by: 
\begin{equation}
  R=\sqrt{3t/ \Omega} 
  \,,
  \label{eq:R_vs_Omega}
\end{equation}
where $\Omega = \sqrt{\omega_e^2 + \omega^2}$ is the total
frequency of oscillation of the electron. Notice that the ground
state energy of \eref{eq:H_WignerCrystal} is $E_0 = -6 t - 3 e^2/(2 r_o)
+ 3   \Omega/2$ and hence the relative gain in free energy is:
\begin{equation}
  \Delta\mathcal{F}_{WP}=-2t+3 \left(\Omega-\omega_{e}\right)/2+4 \pi
  R^{3}T\log\left(2S+1\right)/3.
  \label{eq:DF_WP}
\end{equation}
Minimization of \eref{eq:DF_WP} with respect to $\omega$ gives:
\begin{equation}
  R(T)=\left\{ \begin{array}{c}
  \begin{array}{ll}
  R_S \, \, \left(T^*/T\right)^{1/5} & ,T>T^{*}\\
  R_S & ,T \le T^{*}\end{array}\end{array}\right. 
  \,, 
  \label{eq:Req_WC_2}
\end{equation}
where $R_S = [3t/(  \omega_e)]^{1/2}$ is the saturation radius, and
$T^{*}/t= 9/[4\pi R_S^5 \log\left(2S+1\right)]$ is the temperature below
which the polaron radius saturates due to the interplay between the DEM
and the Coulomb interaction.

\subsection{Wigner Crystal Melting}
\label{sec:WignerCrystalMelting}

It is clear that the previous results are valid for temperatures so low as not
to melt the Wigner solid. The calculations for the independent polaron
model reveal that the temperatures for polaron stability are already
typically small for reasonable values of $t$, but the electronic
solid is much more sensitive to the temperature. The Wigner crystal 
melting temperature, $T_M$, can be estimated from the Lindemann's criteria
\cite{Jones:1996}: $T_M \approx 0.01 (e^2/a) n_e^{1/3}$. 
It is known from several numerical calculations
\cite{Candido:2004,Ortiz:1999,Jones:1996} on the stability of the one component
plasma, that the maximum densities and temperatures at which the Wigner crystal
can exist correspond to $r_{s}\sim50\textrm{-}100$, and $T\sim10\,\textrm{K}$.
Values of $r_{s}\sim50\textrm{-}100$ correspond to
$n_{e}\sim10^{-6}\textrm{-}10^{-5}$ for  $t=1\,\textrm{eV}$ and
$a=4\,\textrm{\AA}$. The region of stability of the polaronic crystal is show
in \fref{fig:PhDiag_WS_Polarons}. Due to the absence of magnetic interactions
between different polarons, the Wigner crystal should be a superparamagnet: the
local moments within the zero point radius of the electrons are expected to
respond collectively. 
In the presence of other long-range interactions (such as dipole-dipole) the
polaronic Wigner crystal can exhibit long range magnetic order. Increasing the
electron density at $T=0$ causes the Wigner crystal to quantum melt at a
critical density with two possible outcomes: a paramagnetic polaronic Fermi
liquid or a fully polarized ferromagnet. In both cases the carriers are mobile
and can screen the long-range part of the Coulomb interaction leading
to a Fermi liquid state.  
At finite temperatures, where the electron state cannot be described by the
zero point motions implicit in \eref{eq:H_WignerCrystal}
alone, the crystal should follow the features of the phase diagram
for the electron gas \cite{Jones:1996}. The characterization of the
system in the neighborhood of the melting point, where the presence
of a polaron liquid is plausible [\fref{fig:PhDiag_WS_Polarons}], 
is restrictively hard, even for the simple electron gas. Far from
this region, where the electron density is high enough to make the
screening process effective, one expects to retrieve the behavior
obtained before within the independent polaron model, and discussed previously.

%
\section{Conclusions}%
\label{sec:Conclusions}

The pure \acl{DEM} Hamiltonian \eqref{eq:DEM-DE-Hamiltonian-2} displays
a rich phase diagram, even at the lowest densities. This is in itself
noteworthy insofar as we did not introduce any additional competing
interactions, such as direct magnetic couplings among the local spins.
In addition to its intrinsic tendency for a \ac{FM} transition at low
temperatures, the low density region
of the phase diagram displays an instability towards phase
separation. This phase separated regime is characterized by \ac{FM} droplets
rich in electrons, embedded in a \ac{PM} background, essentially depleted of
electrons. The consideration of electron-electron interactions shows that,
as expected, the phase separated regime is highly suppressed for meaningfull
magnitudes of the electrostatic correction. This suppression is such that the
effective number of electrons inside each \ac{FM} droplet becomes of the order
of unity. Given that, by construction, our \ac{WS} cells containing the
\ac{FM} droplet are neutral, this situation of nearly one electron per
droplet can be alternatively addressed in terms of non-interacting magnetic
polarons. 

This justifies our initial approach to
address the polaronic phase in the low density \acl{DEM}. Within the \acl{IPM}
developed in \sref{sec:DEM-MagPol-IPM}, below a critical density the PM-FM
transition is mediated by a polaronic phase
(cfr.~\fref{fig:MagPol-PolPhaseDiagSimple}). One consequence of this is that
the Curie temperature of the system is much lower than one would obtain if
based only on a PM-FM mean-field approximation. Moreover, by considering the
model parameters already used in \Ref~\onlinecite{Pereira:2004} to describe
other properties of \ac{EuB6} from a \ac{DE} point of view, we obtain the range
of temperatures $T_m<T<\Tc$ for the presence of the polaronic phase in
agreement with experimental observations, and without additional adjustable
parameters.

Further down in the density scales we estimated the conditions for the
stability of a polaronic Wigner crystal. This phase seems plausible in the
ultra diluted situation where Wigner cristalization of the electron gas is
expected. In this case the zero point motion of the electrons can still
provide enough itinerancy to polarize the neighboring local moments, generating
a crystal of bound magnetic polarons. Nevertheless this regime, although
rather appealing from the theoretical point of view, is certainly difficult to
reach experimentally on account of the reduced temperature and density scales
involved.

The current results complement the ones in \Ref~\onlinecite{Pereira:2004} that
pertain, mostly, to the evolution of the system once the homogeneous \ac{FM}
phases sets in, whereas now we have addressed how the transition from
homogeneous \ac{PM} to the onset of homogeneous \ac{FM} takes place.
We can then interpret the ferromagnetic transition in \ac{EuB6} as being 
precipitated by the merging of magnetic polarons which attain the
percolation threshold close to \Tc. At the same time, these results lend
additional support to the interpretation of the phenomenology of \ac{EuB6} from
a \acl{DE} perspective.

%
\section{Acknowledgments}

We acknowledge many motivating and fruitfull discussions with L. Degiorgi and E.
V. Castro.
V.~M.~Pereira is supported by Funda\c{c}\~{a}o para a Ci\^{e}ncia e a Tecnologia
via SFRH/BPD/27182/2006.
V.~M.~Pereira and J.~M.~B.~Lopes~dos~Santos further acknowledge POCI 2010 via
the grant PTDC/FIS/64404/2006. 

%
\appendix
%

%
\section{Effects of Finite Band Filling on Polaron Stability}%
\label{Appendix:MagPol-FiniteBandFilling} 

In the main discussion of the polaronic physics in the \ac{DEM}, the
simplification is made of considering that the electronic energy is simply
accounted by the energy at the bottom of the band, multiplied by the electron
density \eqref{eq:MagPol-Eel_Bottom}.

In order to account for the finite electronic density, we rely on a \emph{rigid
band} approximation for the DOS. This means that we calculate the electronic
energy for a given density of polarons, $n_p$, assuming the \ac{DOS} in the
conduction band doesn't change significantly \cite{Endnote-10}. 
The free energy per lattice site then becomes
\begin{widetext}
\begin{equation}
  F_{\textrm{Pol}}(R,M,n_p) = 
    -6 t n_p\cos\left(\frac{\pi}{R+1}\right)\nonumber +
    \int^{E_F(n_e - n_p,M)}_{E_b(M)} \rho(\epsilon,M) \epsilon d\epsilon
    + (1 - n_p R^3) T\mathcal{S}(M,S)
  \label{eq:PS-F-Pol-Band}
  \,.
\end{equation}
\end{widetext}
In this approximation, the electronic energy is counted essentially by
transferring electrons from the conduction band to the bound states. The Fermi
energy satisfies $E_F(n_e - n_p,M) > E_b(M)$, reflecting the existence of $n_e -
n_p$ electrons in the band. An important difference relative to the case of the
empty band considered in Sec.~\ref{sec:DEM-MagPol-IPM} follows: since we are
allowing the existence of $n_e - n_p$ states extended throughout the system, the
non-polaronic part can still be ferromagnetic below some temperature. So, in
principle, the magnetic polarons could be embedded in a background with finite
spin polarization, and, therefore, we have to introduce the magnetization of the
background, $M$, as a third variational parameter, together with $R$ and $n_p$.
The minimization of \ref{eq:PS-F-Pol-Band} with respect to these parameters, and
the comparison of the resulting equilibrium free energy with the homogeneous
case \eqref{eq:PS-F-Variational-T0}, produces the phase diagram displayed in
Fig.~\ref{fig:PS-PhDiag-PolaronFinite-1}.
%
\begin{figure}
  \centering
  \subfigure[][]{
    \includegraphics*[width=0.46\textwidth]{%
      Figs/Polaron-PhDiag-Finite_Fill_Correc}%
    \label{fig:PS-PhDiag-PolaronFinite-1}
  }
  \subfigure[][]{
    \includegraphics*[width=0.46\textwidth]{%
      Figs/Polaron-Some_Curves_of_Eq_np_and_M}%
    \label{fig:PS-PhDiag-PolaronFinite-2}
  }
  \caption{
    Polaronic stability relative to homogeneous \ac{PM}/\ac{FM} phases when a
    finite band filling is considered.
    \subref{fig:PS-PhDiag-PolaronFinite-1} Phase diagram.
    \subref{fig:PS-PhDiag-PolaronFinite-2} Each frame shows the equilibrium
    values of $n_p$ and $M$ (order parameters) for total electronic
    concentrations of $n_e=0.0006,\,0.0008,\,0.0014,\,0.0016$. The blue
    (triangles) curve shows the equilibrium $n_p$ that is obtained when only
    paramagnetism is allowed (i.e. constraining $M=0$). 
  }
  \label{fig:PS-PhDiag-PolaronFinite}
\end{figure}
%
The overall qualitative features obtained previously in
Fig.~\ref{fig:MagPol-PolPhaseDiagSimple} remain basically the same, the
important differences now being: (i) The \ac{PM}--\ac{FM} transition curve
(dashed line) now places $T_C$ for homogeneous phases at higher temperatures
than the ones obtained with the de~Gennes treatment; (ii) The polaron stability
temperature, $T_m$, is seen to increase with density if $n_e\lesssim0.001$, just
as expected because, the higher the density, the higher the Fermi energy and the
more favorable it becomes to create a polaron for the same price in entropy;
(iii) The reentrance of the polaronic phase is now very pronounced, being a
consequence of (i). It is nonetheless interesting to observe that the critical
density for the stabilization of the polaronic phase ($n_e\simeq0.004$) and the
typical stabilization temperature, $T_m$, are almost exactly the same as the
ones encountered in Sec.~\ref{sec:DEM-MagPol-IPM}. 

The nature of the transitions as the temperature is lowered is as follows (see
Fig.~\ref{fig:PS-PhDiag-PolaronFinite-2} for reference). Below $n_e\simeq0.001$
the PM--Pol transition is continuous; $n_p(T)$ varies continuously from 0 to the
saturation value $n_p = n_e$, and no \ac{FM} is stabilized except at very low
temperatures, below the Pol phase. For $n_e \gtrsim 0.001$, \ac{FM} is
stabilized with a continuous PM--FM transition; \ac{FM} persists only until
$T_m(n_e)$ is reached, at which point the polarons set in; the FM--Pol
transition is discontinuous because the magnetization drops to zero and $n_p$
jumps from 0 to quasi-saturation upon crossing $T_m$. This can be seen clearly
in the bottom frames of Fig.~\ref{fig:PS-PhDiag-PolaronFinite-2}: above
$n_e\gtrsim 0.001$ the curve $T_m$ jumps discontinuously at $T_m(n_e)$ to meet
$n_p'$. 
At lower temperatures, when the $T_C$ line (red/circles) is crossed, the
magnetization jumps to the value found in an homogeneous FM phase, concurrently
with a discontinuous drop of $n_p$ from $n_e$ to zero. Notice that the $T_C$
line is barely changed by the consideration of a finite band filling. This is
related to the fact that, when $T_C$ is reached, the polaron density has long
since saturated at $n_e$, thus emptying the band from carriers.

%
\section{Finite Size Corrections to the Electronic DOS}%
\label{Appendix:MagPol-FiniteSizeCorrection}

To estimate the $1/L$ corrections to the ground state energy of an electron gas
consider free electrons inside a box of dimensions $L_x, L_y$ and $L_z$
\cite{Pathria:1996}. The
electronic spectrum is
\begin{equation}
  E_k = \frac{\hbar^2}{2m} k^2
  \,, \qquad \text{with} \quad
  \vec{k} = \pi \left( \frac{n_x}{L_x}, \frac{n_y}{L_y}, \frac{n_z}{L_z} \right)
  \,, \quad (n_i \ge 1)
  \label{eq:Appendix-Spectrum}
  \,.
\end{equation}
The integrated \ac{DOS} will clearly be 
\begin{equation}
  \Omega(E) = \sum_{n_x,n_y,n_z \ge 1} \Theta\left(\kappa^2 - k(n_x,n_y,n_z)^2
\right)
  \label{eq:Appendix-Omega}
  \,,
\end{equation}
which corresponds, geometrically, to the set of integers $(n_x,n_y,n_z)$ bounded
by the ellipsoid
\begin{equation}
  \frac{\kappa^2}{\pi^2} \ge 
  \frac{n_x^2}{L_x^2} + \frac{n_y^2}{L_y^2} + \frac{n_z^2}{L_z^2}
  \label{eq:Appendix-Elipsoid}
  \,.
\end{equation}

\begin{figure}
  \centering
    \includegraphics*[width=0.8\columnwidth]{%
      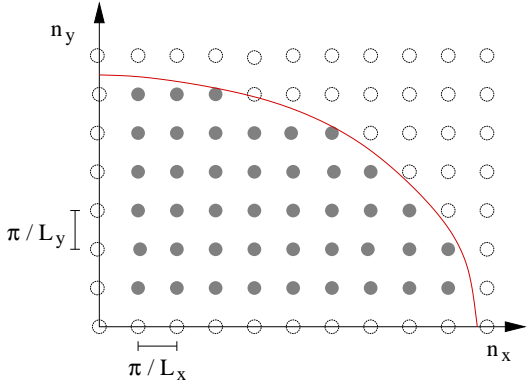}
  \caption{
    Discrete Phase Space.
  }
  \label{fig:Appendix-PhaseSpace}
\end{figure}

In the thermodynamic limit, $L_{x,y,z}\to\infty$ and the usual procedure
consists in replacing the discrete number of states satisfying
\eqref{eq:Appendix-Elipsoid} by the  volume of the ellipsoid in the first
octant, divided by the elementary phase space volume. It is obvious that, by
doing this, one is either neglecting or overcounting some of the points in phase
space that rightfully satisfy \eqref{eq:Appendix-Elipsoid}. This happens mainly
at the \emph{boundaries}, and is all right when $L_{x,y,z}\to\infty$ because the
errors are of the order of $1/L$ or $1/L^2$
(Fig.~\ref{fig:Appendix-PhaseSpace}). However, when $L$ is finite, such
corrections are clearly relevant. A particularly important one arises from the
fact that, when we calculate the volume of the ellipsoid, the points lying at
the coordinate axes are automatically included, whereas from
\eqref{eq:Appendix-Spectrum} they should not be. So let
\begin{equation}
  \Omega'(E) = \sum_{n_x,n_y,n_z} \Theta\left(\kappa^2 - k(n_x,n_y,n_z)^2
                \right)
  \,,
\end{equation}
with $n_i \in \mathbb{Z}$, which relates to $\Omega(E)$ defined in
\eqref{eq:Appendix-Omega} by
\begin{align}
  \Omega'(E) &= 8 \, \Omega(E) 
    + \sum_{n_x, n_y} \Theta\left(\kappa^2 - k(n_x,n_y,0)^2 \right) 
    \nonumber\\
    & + \sum_{n_x, n_z} \Theta\left(\kappa^2 - k(n_x,0,n_z)^2 \right)
    \nonumber\\
    & + \sum_{n_y, n_z} \Theta\left(\kappa^2 - k(0,n_y,n_z)^2 \right)
    \nonumber\\
    & - \sum_{n_x}      \Theta\left(\kappa^2 - k(n_x, 0, 0)^2 \right)
    \nonumber\\
    & - \sum_{n_y}      \Theta\left(\kappa^2 - k(0, n_y, 0)^2 \right)
    \nonumber\\
    &- \sum_{n_z}      \Theta\left(\kappa^2 - k(0, 0, n_z)^2 \right)
    + 1
  \label{eq:Appendix-OmegaPrime}
  \,.
\end{align}
The above result reflects the fact that, when calculating the continuum volume
enclosed by the ellipsoid, one is adding an \emph{extra} portion of phase space
near the coordinate axes that should not be included, as in
Fig.~\ref{fig:Appendix-PhaseSpace}. Assuming that $\kappa^2 / \pi^2$ is still
reasonably greater than $L_x^{-1},L_z^{-1} \text{ or } L_z^{-1}$, these terms
correspond to 
\begin{gather}
  \Omega'(E)  \simeq \text{Vol. ellipsoid with axes } L_x,L_y,L_z \,;
\nonumber\\ 
  \sum_{n_i, n_j} \Theta\left(\kappa^2 - k(n_i,n_j,0)^2 \right) 
     \simeq \text{ Area of ellipse of axes } L_i, L_j \,; \nonumber\\
  \sum_{n_i} \Theta\left(\kappa^2 - k(n_i,0,0)^2 \right) 
     \simeq \text{ Length of the axis } L_i \,. 
\end{gather}
Therefore, we have that
\begin{widetext}
\begin{equation}
  \frac{4}{3}\pi \left( \frac{k}{\pi} \right)^3 L_x L_y L_z 
    = 8 \Omega(E) + \pi \left( \frac{k}{\pi} \right)^2 (L_x L_y + L_y L_z + L_z
L_x)
    - 2 \frac{2}{\pi} (L_x + L_y + L_z)
    + 1
  \,,
\end{equation}
implying that the corrected phase space volume is actually
\begin{equation}
  \Omega(E) = \frac{\kappa^3}{6\pi^2} L_x L_y L_z 
    - \frac{\kappa^2}{8\pi} (L_x L_y + L_y L_z + L_z L_x)
    + \frac{\kappa}{4\pi} (L_x + L_y + L_z) 
    - \frac{1}{8}  
  \label{eq:Appendix-PhSpaceVolumeCorrection}
  \,.
\end{equation}
\end{widetext}
There is still the error associated with the under/over estimates of the volume
near the surface of the ellipsoid. This gives an additional contribution of the
order $\mathcal O(\kappa)$ \cite{Endnote-11}. But since the precise numerical
factors are impossible to extract
analytically, we do not include this correction. Consequently, the expression
above is meaningful only down to $\mathcal O(\kappa^2)$. Given that the volume
and surface area of the volume enclosing the electron gas are related to the
$L$'s by $L_x L_y L_z = V$ and $L_x Ly + L_y L_z + L_z L_x = S /2$, the above
can be summarized as 
\begin{equation}
  \Omega(\kappa) \simeq \frac{\kappa^3}{6 \pi^2} V - \frac{\kappa^2}{16 \pi} S
  \,, 
\end{equation}
from which the corrected \ac{DOS} follows:
\begin{equation}
  g(\kappa) = \frac{\partial\Omega(\kappa)}{\partial \kappa} 
            = \frac{k^2}{2\pi} V - \frac{k}{8\pi} S
            = g_\infty(\kappa) \left( 1 - \frac{\pi}{4\kappa}\frac{S}{V} \right)
  \label{eq:Appendix-PhSpaceCorrectedDOS}
  \,.
\end{equation}
For instance, taking the leading correction for electrons inside a cubic box
($L_i = L$), implies a correction to the ground state energy per electron
reading
\begin{equation}
  \frac{E}{N_e} = \frac{E_\infty}{N_e} 
    \left[1 + \frac{15}{8}\frac{\pi}{L} \left(\frac{1}{6\,\pi^2
n}\right)^{1/3}\right]
  \,, 
  \label{eq:Appendix-PhSpaceCorrectedEnergy-Cube}
\end{equation}
with
\begin{equation}
  \frac{E_\infty}{N_e} = \frac{3}{5} \frac{\hbar^2}{2m} \kappa_{F_\infty}^2 
  \,.
\end{equation}
In the main text, we are interested in the corrections to the energy of an
electron gas confined to finite sized spherical bubbles. We notice that for a
cube of side $L$, 
\begin{equation}
  \left(\frac{S}{V}\right\vert_\text{cube} = 
    \left(\frac{S}{V}\right\vert_\text{inscribed sphere} 
  \,,
\end{equation}
the surface to volume ratio, equals the same ratio for the inscribed sphere with
$R=L/2$.
Hence we take the result \eqref{eq:Appendix-PhSpaceCorrectedEnergy-Cube} and
simply substitute $L$ by $2R$, obtaining
\begin{equation}
  \frac{E}{N_e} = \frac{E_\infty}{N_e} 
    \left[1 + \frac{15}{16} \left(\frac{\pi}{6\, n}\right)^{1/3} \frac{1}{R}
\right]
  \label{eq:Appendix-PhSpaceCorrectedEnergy-Sphere}
  \,,
\end{equation}
which provides an estimate of the leading corrections to the ground state (and
$T=0$) energy of the confined electron gas.

%
%


\begin{acronym}[MOSFET]
  \acro{AFM}{antiferromagnetic}
  \acro{BMP}{bound magnetic polaron}
  \acro{BZ}{Brillouin zone}
  \acro{CMR}{colossal magnetoresistance}
  \acro{DEM}{double exchange model}
  \acro{DE}{double exchange}
  \acro{DMS}{diluted magnetic semiconductors}
  \acro{DOS}{density of states}
  \acro{EuB6}[$\mathrm{EuB_6}$]{europium hexaboride}
  \acro{FMP}{free magnetic polaron}
  \acro{FM}{ferromagnetic}
  \acro{HTA}{hybrid thermodynamic approach}
  \acro{IPM}{independent polaron model}
  \acro{KLM}{Kondo lattice model}
  \acro{MC}{Monte Carlo}
  \acro{PM}{paramagnetic}
  \acro{PS}{phase separation}
  \acro{WS}{Wigner-Seitz}
\end{acronym}

\bibliographystyle{apsrev}
\bibliography{magpol-phsep}

\begin{thebibliography}{76}
\bibitem[{End({\natexlab{d}})}]{Endnote-6}
\bibinfo{note}{In fact, we can do the Maxwell construction in both the
  $\mathcal F_{eq}(n)$ or the $\mu(n)$ diagram. In the latter, the Maxwell
  construction corresponds to finding the horizontal line that yields exactly
  $A_1=A_2$ \cite{Callen:1985}.}
\bibitem[{End({\natexlab{f}})}]{Endnote-8}
\bibinfo{note}{At first sight it might seem that the electrons are
  disappearing! In reality it just means that, according to the \ac{WS}
  picture, more and more \ac{WS} cells appear in the system, as $R$ diminishes,
  because the total number of electrons and the total volume are fixed.}
\bibitem[{End({\natexlab{g}})}]{Endnote-9}
\bibinfo{note}{In order to compare directly with the polaronic stability
  region, the Langevin entropy was replaced by the Brillouin entropy in the
  magnetic contribution to the free energy, for consistency. The behavior is
  the same irrespective of which case is considered, up to global numerical
  factors.}
\bibitem[{End({\natexlab{h}})}]{Endnote-10}
\bibinfo{note}{We are actually allowed to do that since only a few bound states
  appear beyond the band edge, as can be seen in
  Fig.~\ref{fig:MagPol-DOS-and-Spectrum}, and the profile of the bulk \ac{DOS}
  is not altered.}
\bibitem[{End({\natexlab{i}})}]{Endnote-11}
\bibinfo{note}{This can be inspected numerically by plotting the difference
  between \eqref{eq:Appendix-PhSpaceVolumeCorrection} and
  \eqref{eq:Appendix-Omega} against $\kappa$. For $L_x=L_y=L_z=L$ the
  difference appears majored by a term linear in $\kappa$.}
\end{thebibliography}

\end{document}